\documentclass[apj]{emulateapj}
\usepackage{url,amsfonts,epsfig}
\usepackage{graphicx}
\usepackage{amsmath}
\usepackage{amssymb}
\usepackage[section]{placeins}
\usepackage{chngpage}
\usepackage{calc}
\usepackage{subfigure}
\usepackage{mathtools}
\usepackage{color}

\usepackage{natbib}
\makeatletter 
\renewcommand\@biblabel[1]{}

\shorttitle{stellar remnants around massive black holes}
\shortauthors{Antonini}
\begin{document}
\def\gap{\;\rlap{\lower 2.5pt
\hbox{$\sim$}}\raise 1.5pt\hbox{$>$}\;}
\def\lap{\;\rlap{\lower 2.5pt
 \hbox{$\sim$}}\raise 1.5pt\hbox{$<$}\;}

\title{On the distribution of stellar remnants around massive black holes: slow mass segregation, star cluster
inspirals and correlated orbits}

\author{Fabio Antonini}
\email{antonini@cita.utoronto.ca}
\affil{Canadian Institute for Theoretical Astrophysics, University of Toronto,
60 St. George Street, Toronto, Ontario M5S 3H8, Canada}

\begin{abstract}
We use $N$-body simulations as well as analytical techniques to study the long term  dynamical evolution of stellar black holes (BHs) at the Galactic center~(GC) and to put constraints on their number and  mass distribution.
Starting from  models that have not yet achieved a state of collisional equilibrium, we find that time scales associated with cusp regrowth can be longer than the Hubble time.
Our results cast doubts on standard models that postulate high densities of BHs near the GC and motivate studies that start from initial conditions which correspond to well-defined physical models.  
For the first time, we consider the distribution of BHs in a dissipationless  model for the  formation of the Milky Way nuclear cluster~(NC), in which massive stellar clusters merge to form a compact nucleus.
We simulate the consecutive merger of $\sim 10$ clusters containing an inner dense sub-cluster of BHs.
After the formed NC is evolved for $\sim 5$~Gyr, the BHs do form a steep central cusp, while the stellar distribution maintains properties that resemble those  of the GC NC. 
Finally, we  investigate the effect of BH perturbations on the motion of the GC S-stars, as a means of constraining  the number  of the perturbers. We find that reproducing the quasi-thermal character of the S-star orbital eccentricities requires $\gap 1000$ BHs within $0.1~$pc of Sgr~A*. A dissipationless formation scenario for the GC NC is consistent with this lower limit and therefore could reconcile the need for high central densities of BHs (to explain the S-stars orbits), with the ``missing-cusp'' problem of the GC giant star population.
  \end{abstract}
 

\section{introduction}

Massive nuclear clusters~(NCs) are observed at the center of many galaxies, over the whole Hubble sequence. 
The frequency of nucleation among galaxies less luminous than
$\sim10^{10.5}~L_\odot$ is close to $90~\%$  as determined by ACS HST  observations of
galaxies in the Virgo and  Fornax galaxy clusters~\citep{carollo,BSM:02,Cote,turner+12}.
The study of NCs is of great interest for our understanding of
galaxy formation and evolution as indicated by the fact that a number of fairly tight correlations are observed 
 between their masses  and  global 
properties of their host galaxies such as  velocity dispersion and bulge mass~\citep{F:06,WH:06,GS:09,SG12,Nathan}.
Intriguingly, similar scaling  relations are obeyed by  massive
black holes~(MBHs) which are predominantly found in massive galaxies that, however, show 
little evidence of nucleation~\citep[e.g.,][]{GS:09,NW12}. 
The existence of such correlations  might indicate a direct link 
among large galactic spacial  scales and the much smaller scale of the nuclear environment, and suggests that
NCs contain information about the processes that have shaped the central regions of 
their host galaxies.

How NC  formation takes place at the center of galaxies is still largely 
debated~\citep[e.g.,][]{Hartmann,gnedin+13,carlberg+hartwick14,mpl14}. 
Relatively recent work has shown that ``dissipationless''  models can reproduce without obvious difficulties
the observed properties~\citep{turner+12} and scaling relations \citep{me13} of NCs. 
In these models a NC forms through the inspiral of massive stellar clusters into the center
due to dynamical friction where they merge to form a compact nucleus~\citep[e.g.,][]{TOS,CDM08,capuzzo93}.
Alternatively, NCs could have formed locally as a result of radial gas inflow into the galactic center
accompanied  by efficient dissipative processes~\citep{Sch08,Milos04}.  
Naturally,  dissipative and dissipationless processes are not exclusive and both could be important for the formation and evolution 
of NCs~\citep{Hartmann,AM12-2,de-lorenzi+13}.

The Milky Way NC, being only $8~$kpc away, is currently the only NC 
 that can be resolved in individual stars and for which a  kinematical structure and 
 density profile can be reliably determined~\citep{gen+10}. This
 offers the unique possibility to resolve the stellar population, to study the composition and dynamics 
 close to a MBH and put constrains on different NC formation scenarios.
The Milky Way NC has an estimated mass of $\sim 10^{7}M_{\odot}$ \citep{LZM,S09}, 
and it hosts a massive black hole  of $\sim 4\times 10^6 M_\odot$~\citep{Genzel03,Ghez08,Gil}
whose gravitational potential dominates over the stellar cusp potential 
out to a radius of roughly $3$pc - the MBH radius of influence. 
A handful of other galaxies are also known to contain both a NC and a MBH, which typically
have comparable masses~\citep{seth}. Population synthesis models suggest 
that roughly $80\%$ of the stellar mass in the inner parsec of the Milky Way is in ($>5~$Gyr) old stars~\citep{Ol11}
although the light is dominated by the young stars. This appears to be typically the case also in
most NCs observed in external galaxies \citep{Rossa}.

Over the last decades observations of the Galactic NC have  led  
to a number of puzzling discoveries. These
include: the presence of a young population of stars (the S-stars) near Sgr~A* in  an 
environment extremely hostile to star formation~\citep[paradox of youth,][]{Morriss1993,sch02};
and a significant paucity of red giant stars in the inner half a  parsec~\citep[conundrum of old age,][]{M:10}.
Number counts of the giant stars at the Galactic center~(GC) show that their
visible distribution is in fact quite inconsistent with
the distribution of stars expected for a dynamically relaxed population  near a dominating Keplerian
potential~\citep{BSE,D:09,B:10}: instead of a steeply rising \citet{BW:76} cusp, there is a $\sim 0.5$ pc core. 
The lack of a Bahcall-Wolf cusp in the giant distribution  casts doubts on  dynamical 
relaxed, quasi-steady-state models  of the GC
which postulate a  high central density of stars and stellar black holes~(BHs).
In these models the central distribution  of stars and BHs is determined by just a handful of  
parameters: the MBH mass; the total density outside the relaxed region; the slope of the initial mass function~\citep[IMF,][]{Merritt-book}. Given the unrelaxed form of the density profile of stars,
making predictions about  the distribution of the stellar remnants becomes
a much more challenging, time-dependent, problem 
 susceptible  to the initial conditions and to the (yet largely unconstrained) formation process 
of the NC~\citep{AM12}.  


%

Understanding the distribution of
the ``stellar remnants'' in systems similar to the Milky Way's NC  is crucial
in many respects. Examples include randomization of the S-star orbits via gravitational encounters~\citep{Perets2009},
 warping of the young stellar disk~\citep{kocsis+tremaine11}, and 
 formation of X-ray binaries~\citep{muno+05}. Stellar nuclei similar to that of the Milky Way
are also the location of astrophysical processes 
that are potential  gravitational wave~(GW) sources both for   ground and space  based 
laser interferometers. These include the merger of 
compact object binaries  near MBHs~\citep{me+perets12}, and
the capture of BHs by MBHs,  called ``extreme mass-ratio inspirals"~\citep[EMRIs][]{AS12}.
The efficiency of these dynamical processes and  rate estimates for GW sources 
are very sensitive to the number of BHs 
near  the center. Therefore, a fundamental question  is whether given a prediction for the initial distribution of stars
and BHs, the system is old enough  that the heavy remnants  had time to relax and 
segregate to the  center of the Galaxy.

 

Motivated by the above arguments,   we consider
  the long-term evolution of BH populations at the center of galaxies, starting from 
  different assumptions regarding their initial distribution. Since the stellar BHs at the GC are not directly detected, time-dependent numerical  calculations, like the ones presented below, are crucial
 for understanding and making predictions about the distribution  of stellar remnants at the center of galaxies.  
  
In Section~\ref{sec-mass-seg} we explore the evolution of  models 
in which stars  and BHs follow initially  the same spatial distribution which is far from being
in collisional equilibrium. Contrary to some previous claims~\citep{preto-amaro10}, we find that in these models the time to regrow a cusp 
in both the BH and the star distribution  is longer than the age of the Galaxy.
For realistic number fractions of BHs, our simulations demonstrate that over the age of the Galaxy 
the presence of a heavy component has little effect on the evolution of the stellar component.

In Sections~\ref{timescales},~\ref{ics} and ~\ref{results} we discuss the evolution of BHs 
in a globular cluster merger model for NCs.
We present the results of  direct $N$-body simulations  
of the merger of  globular clusters containing two mass populations: stars and BHs. 
These systems were   in an initial   state of mass segregation with the BH population concentrated toward the cluster core.
Each cluster was  placed on a circular orbit with galactocentric radius of $20~$pc in a $N$-body system 
containing a central MBH. 
We find that the inspiral of massive globular clusters in the center of the Galaxy 
constitutes an efficient source term of BHs in these regions. After about ten inspiral events 
the BHs are highly segregated to the center. After
a small fraction of the nucleus relaxation time (as defined by the main stellar population)
the BHs attain a nearly-steady state distribution; at the same time 
the stellar density profile exhibits a $\sim 0.2$pc core, 
similar to the  size of  the core in the distribution of  stars at the GC. 
 Our results indicate that
standard models, which assume the same initial phase space distribution
for BHs and stars, can lead to misleading  results regarding the current dynamical state of the Galactic center.

We discuss the implications of our results in Section~\ref{discussion}. In particular, we show
that in order to reproduce the quasi-thermal form of the observed eccentricity distribution of the S-star orbits,
about $1000~$BHs should be present inside $\sim 0.1$pc of Sgr~A*. This number  
appears to be consistent with the number of BHs  expected in a model 
in which the Milky Way NC formed trough the orbital decay and merger of about $10$ 
massive clusters.

Our main results are summarized in Section $7$.


  \section{slow mass segregation at the Galactic center}\label{sec-mass-seg}
 In this section we study the long term dynamical evolution of multi-mass models for the Milky Way NC.
The primary goal of this study is to understand the evolution of the  distribution 
of stars and BHs over a time of order the central relaxation time of the nucleus, 
 starting from initial conditions that are far from being in collisional equilibrium.

\begin{figure}
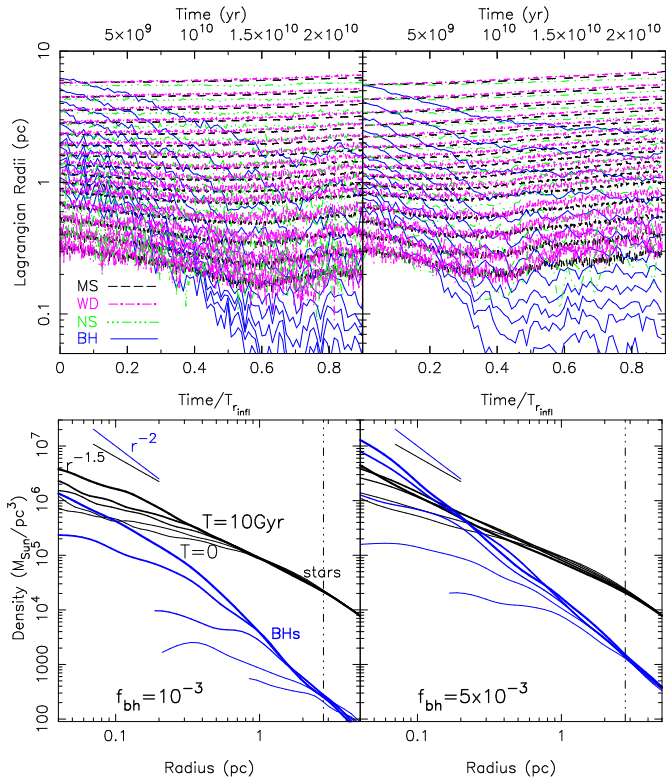

\centering
~\includegraphics[width=0.303\textwidth,angle=270]{Figure1a.ps} \\
  \includegraphics[width=0.265\textwidth,angle=270]{Figure1b.ps} 
  \caption{Top panels show the  Lagrangian radii for the four
  stellar species during the $N$-body simulations with BH number fractions: $f_{\rm bh}=10^{-3}$~(left panels) and
 $f_{\rm  bh}=5\times10^{-3}$~(right panels). 
  Top tick-marks give times after scaling  to the Milky Way; we adopted a relaxation time at 
  the Sgr~A* influence radius of $25$~Gyr~\citep[e.g.,][]{M:10}. Bottom panels show 
 the density profile of stars and BHs at $t=(0,~2.5,~5,~7.5,~10)$Gyr; central density increases with time. 
 Clearly, even after a time of order $10~$Gyr, the  distribution of stars and BHs in 
 our models can be very different from the relaxed multi-mass
 models that are often used to describe  the center of galaxies. Vertical line marks the MBH influence radius.
  }\label{mass-seg}
\end{figure}

\begin{figure*}
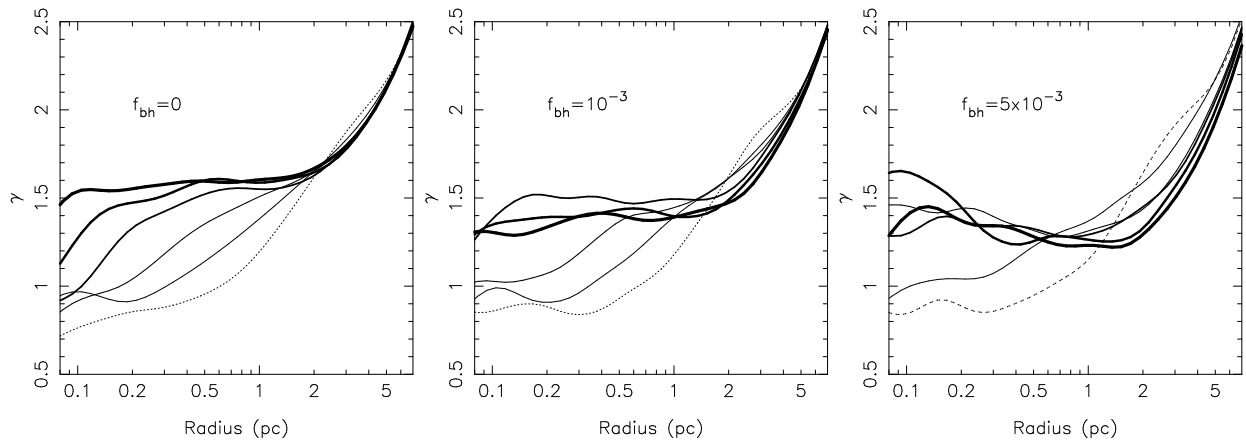
\centering
\includegraphics[width=0.3\textwidth,angle=0]{Figure2a.ps} 
  \includegraphics[width=0.3\textwidth,angle=0]{Figure2b.ps} 
  	  \includegraphics[width=0.3\textwidth,angle=0]{Figure2c.ps} 
  \caption{Evolution of the density slope, $\gamma \equiv -d \log \rho/d \log r$, of the main-sequence density
  profile in  multi mass $N-$body models~(central and right panels), compared to a model with only one mass component~(left panel). The continue curves show profiles at $(0.2,~0.4,~0.6,~0.8,~1)\times T_{r_{\rm infl}}$; increasing line width corresponds to increasing time. The dashed curve corresponds to the initial model.
 Adding a BH component accelerates the growth of a density cusp in the stellar component.  However
 the time to regrow a cusp in these models is always longer than $0.2T_{r_{\rm infl}}$, i.e., $ {\rm 5Gyr}$
when scaled to the Milky Way, a time longer than the mean stellar age of the Galactic NC.}  \label{mass-seg-slp}
\end{figure*}

\subsection{ Evolution toward the steady state}
 We consider four mass groups representing main sequence stars (MSs), white dwarfs~(WDs), neutron stars~(NSs) and BHs.  After the quasi steady-state is attained, the stars are expected to follow a central $r^{-3/2}$ cusp, while the heavier 
 particles will have a steeper $r^{-2}$ density profile~\citep[e.g.,][]{AX:05}.
 We assume that all species have the same phase space distribution initially as it would
 be expected for a violently relaxed system. This is the assumption that was made in most previous 
 papers~\citep[e.g.,][]{F06,HA06,M:10}.
 We specify the mass ratio, $m_{\rm wd}/m_{\star}=0.6$, $m_{\rm ns}/m_{\star}=1.4$, $m_{\rm bh}/m_{\star}=10$,
 between the mass group particles and respective number fractions,
$f_{\rm wd}=N_{\rm wd}/N_{\star}$, $f_{\rm ns}=N_{\rm ns}/N_{\star}$, $f_{\rm bh}=N_{\rm bh}/N_{\star}$. 

 Number counts of the old stellar population at the GC are consistent with 
 a density profile of stars that is flat or slowly rising toward the MBH inside its sphere of influence  and within a radius of
 roughly $\sim 0.5~$pc \citep{BSE,D:09,B:10}. Outside this radius the density falls off as $r^{-2}$. 
\citet{M:10} showed that a core of size  $\sim 0.5~$pc is a natural consequence of two-body relaxation acting over 
 $10~$Gyr, starting from a core of radius $\sim 1~$pc.
 It is therefore of interest to study the evolution of the BH distribution for a time of order the age of the Galaxy 
 and starting from a density distribution with a parsec-scale core.
 We adopt the truncated broken-power-law model:
 \begin{equation}
\label{den}
\rho(r)=\rho_0 \left(  \frac{r}{r_0} \right)^{-\gamma_i}
 \left[
1+\left( \frac{r}{r_0} \right)^{\alpha}\right]^{(\gamma_i-\gamma_{e})/{\alpha}}\zeta(r/r_{\rm cut})~,
\end{equation}
were  $\zeta(x)=\frac{2}{\mathrm{sech}(x)+\mathrm{cosh}(x)}$,
 $\alpha$ is a parameter that defines the transition strength between inner and outer power laws, $r_0$
 is the scale radius and $r_{\rm cut}$ is the truncation radius of the model. The values adopted 
 for these parameters were: $r_0=1.5$pc, $\alpha=4$, $\gamma_e=1.8$ and $r_{\rm cut}=6$pc.
 We included a central  MBH of mass~$M_{\bullet}=4\times 10^6~M_{\odot}$ and
generated the  models $N$-body representations via numerically calculated distribution functions. 
 The central slope was set to $\gamma_i=0.6$, the smallest density slope index consistent with an isotropic 
 distribution for the adopted density model and potential. 
  
 The normalizing factor $\rho_0$ was chosen in such a way that the corresponding density profile reproduces 
 the coreless density model:
 \begin{equation}
 \rho(r)=1.5\times 10^5 \left( \frac{r}{1{\rm pc}} \right)^{-1.8} M_{\odot} {\rm pc}^{-3}
 \end{equation}
 outside the core. This choice of normalizing constant gives a mass density at $1~$pc similar to what 
 it is inferred from observations~\citep[e.g.,][]{okf}, and gives a total mass in stars 
 within this radius of $\sim 1.6 \times 10^6~M_{\odot}$.
The fact that our  models are directly scalable to the observed stellar density distribution of 
 stars at the GC is important if we want to draw conclusions about the current dynamical state
 of stars and BHs at the GC. We note, for example, that the merger  models of 
 \citet{GM12}  had core radii that were substantially larger than the MBH influence radius. 
 As also noted by these authors, this  simple fact precluded a unique scaling of  their models 
 to the Milky Way -- at least in the Galaxy's current state 
in which  the stellar core size ($\sim 0.5~$pc) is much smaller than the Sgr~A* influence radius~($\sim 3~$pc). 
 
 We run three simulations with  $N=132$k particles.  These simulations differ
with each other by the adopted number fractions of the four mass groups:  (i) $f_{\rm wd}=f_{\rm ns}=f_{\rm bh}=0$;
(ii)  $f_{\rm wd}=10^{-1}$, $f_{\rm ns}=10^{-2}$, $f_{\rm bh}=10^{-3}$;
(iii) $f_{\rm wd}=2\times10^{-1}$, $f_{\rm ns}=2\times10^{-2}$, $f_{\rm bh}=5\times10^{-3}$.
The latter two set of values correspond roughly to the number fractions
expected from a standard and from a top-heavy IMF respectively.
 A fraction $f_{\rm bh}=10^{-3}$ is what expected for a  standard (Kroupa-like) IMF
and it is the value typically adopted in previous studies~\citep[e.g.][]{HA05,HA06}.
Although a larger fraction of stellar remnants might be possible, for instance if the Galactic center always obeyed a top heavy initial mass function, 
the observationally constrained  mass-to-light ratio of the inner parsec  limits the BH fraction to only a few percent and it is
more consistent with a ratio and a total mass of BHs predicted by a standard IMF~\citep{lockmann+10}.
We evolved these systems for a time equal to the relaxation time, $T_{r_{\rm infl}}$, computed
at the sphere of influence of the MBH. 
The relaxation time was evaluated  using the expression~\citep{Spitzer}:
\begin{equation}
T_{r}=\frac{0.34\sigma(r)^3}{G^2 <m> {\rm ln} \Lambda \rho(r)}
\end{equation}
with $\rho$ the total local mass density, and $<m>$ the average particle mass.
For the Coulomb logarithm we used $\ln \Lambda=\ln \left( r_{\rm infl} \sigma^2 / 2Gm_{\star} \right) \approx 10$,
with $\sigma$ the 1d velocity dispersion outside $r_{\rm infl}=GM_{\bullet}/\sigma^2$.

To scale the $N-$body time length to the Milky Way we consider 
that the relaxation time at the influence radius 
of Sgr~A*, $r_{\rm infl}\approx 3$pc, is $T_{r_{\rm infl}}\approx 25~$Gyr, assuming a stellar mass of 
$1M_{\odot}$~\citep{M:10,AM12}. 
Thus, when scaling  to the GC, a time of $0.4 T_{r_{\rm infl}}$
corresponds to roughly $10~$Gyr.

\begin{figure*}
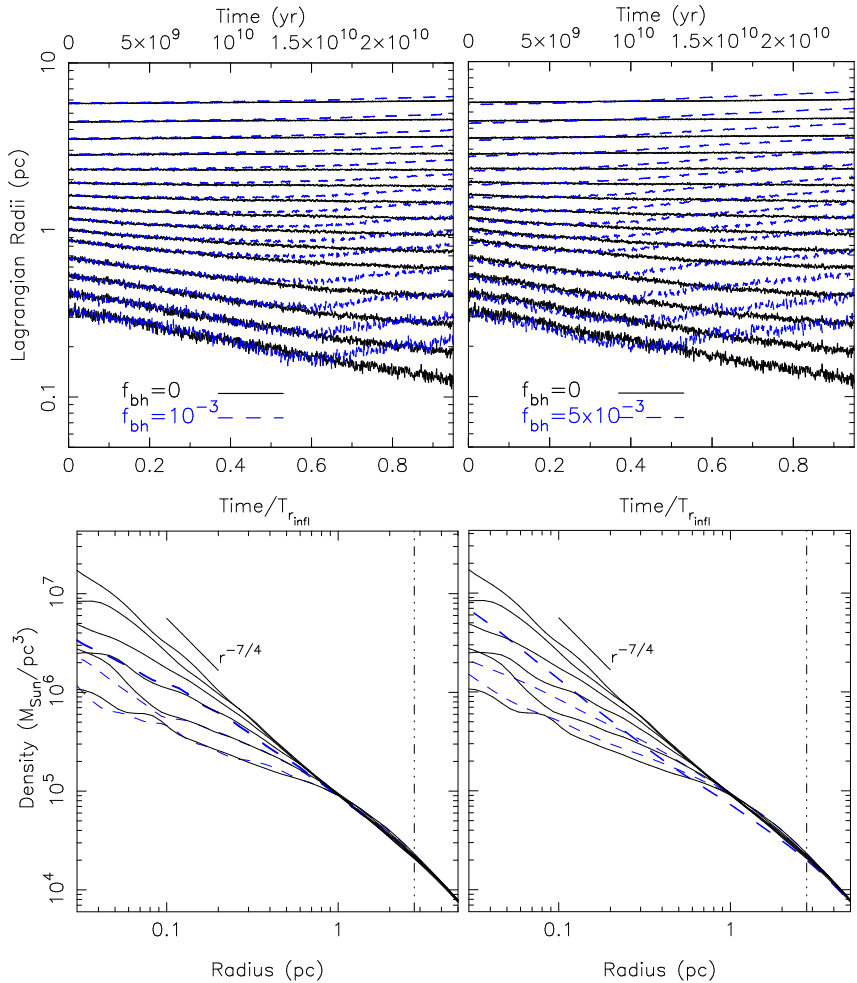

\centering
\includegraphics[width=0.33\textwidth,angle=0]{Figure3a.ps} 
\includegraphics[width=0.29\textwidth,angle=0]{Figure3c.ps}  
 \\ 
\includegraphics[width=0.33\textwidth,angle=0]{Figure3b.ps} 
\includegraphics[width=0.283\textwidth,angle=0]{Figure3d.ps}  
\\
  \caption{  Evolution of the $ f_{\rm bh}=0$ model Lagrangian radii and density profile
  compared to the evolution of models with $ f_{\rm bh}=10^{-3}$ (left panel) and 
  $ f_{\rm bh}=5\times10^{-3}$~(right panel) .
  In the bottom panels the density profile is plotted at $t=(0,~5,~10,~15,~20,~25)$Gyr of evolution.  
  The blue-dashed curves give the evolution of the stellar distribution Lagrangian radii in the models with BHs
  and the bottom panels show the respective stellar density profiles  at  $t=(0,~5,~10)$Gyr. Over a time of order $10~$Gyr 
the evolution of the density profile of stars in the $ f_{\rm bh}=10^{-3}$
 model is not much affected by the presence of the BHs. In the model with $f_{\rm bh}=5\times10^{-3}$ the evolution
 toward the steady state is faster 
 and after $10$Gyr the stars have formed a cusp. Vertical lines give the MBH influence radius.}\label{mass-seg2}
\end{figure*}

  We evolved the initial conditions with the direct $N-$body
 integrator $\phi$GRAPEch \citep{HGMM}. The code implements  a fourth-order Hermite
  integrator with a predictor-corrector scheme
and hierarchical time stepping.
The code combines hardware-accelerated
computation of pairwise interparticle forces 
\citep[using the \textsl{Sapporo} library which emulates the GRAPE interface utilizing GPU boards,][]{sap}
with a high-accuracy chain regularization algorithm to follow the dynamical interactions of 
field particles with the central MBH particle. 
The chain radius was set to $10^{-2}$pc and we used a softening $\epsilon =10^{-6}~$pc. 
The relative error  in total energy was typically $\sim 10^{-4}$ for the accuracy parameter $\eta=0.01$.

Figure~\ref{mass-seg} shows the evolution of the $N$-body models over one relaxation time. 
The heavy particles segregate to the center owing dynamical friction. After
the central mass density of BHs becomes comparable to the density in the other species, the evolution
of the BH population starts being dominated by  BH-BH  self interactions; at the same time
the lighter species evolve in response to dynamical heating from 
the BHs, which causes the local stellar 
densities to decrease and Lagrangian radii to expand. As shown below,
the same heating  rapidly converts the initial density profile into a steeply rising density cusp with slope, 
$\gamma \equiv -d \log \rho/d \log r \approx 3/2$. 
The inclusion of a BH population has therefore two effects on the main sequence population: it lowers the
stellar densities and at the same time
 it accelerates the evolution of  the density of stars toward the $\gamma=3/2$ steady-state form.

  The lower panels of Figure~\ref{mass-seg} display 
 the density profile of stars and BHs over $10~$Gyr of evolution. These plots  show that, starting with a fraction of BHs 
 that corresponds to a  standard IMF:
 (1) after $\sim 10$Gyr
 the density of BHs can remain well below the density of stars at all radii; (2) even after
 10~Gyr of evolution, the density distribution of stars
looks very different  from what  expected for a dynamically relaxed population 
 around a MBH.  
 These findings are in agreement with  the Fokker-Plank simulations of \citet{M:10} 
 but   in  contrast with more recent claims that mass segregation can
rebuild a stellar  cusp in a relatively small fraction of the Hubble
time~(e.g.,  \cite{preto-amaro10}, and the Introduction of \cite{amaro-xian13}). 
Figure~\ref{mass-seg-slp} displays the evolution  of the radial profile of the density profile slope.
Comparing the evolution observed in  models with and without BHs we see that
a cusp in the main-sequence population develops earlier in models with BHs. Figure~\ref{mass-seg-slp} shows  that 
for $ f_{\rm bh}=10^{-3}$ and $ f_{\rm bh}=5\times 10^{-3}$  
a stellar cusp only develops after $\sim 0.6T_{r_{\rm infl}}$ and $\sim 0.4T_{r_{\rm infl}}$ respectively. 
Therefore over the timescales ($\lesssim10~$Gyr) and radii~($r\gtrsim 0.01$pc) of relevance, 
the inclusion of a BH population has little or 
even no influence on the evolution of the lighter populations. 
This latter point is more clearly demonstrated in Figure~\ref{mass-seg2} which directly compares  the Lagrangian radii
 evolution of our $f_{\rm bh}=0$ model  with models with BHs.  The stellar
 populations evolve similarly in these models independently on $f_{\rm bh}$
 until approximately $0.6 $ and  $0.4 \times T_{r_{\rm infl}}$
 for $ f_{\rm bh}=10^{-3}$ and $5\times 10^{-3}$ respectively.
After this time,  heating of the lighter species by the heavy particles starts becoming important
causing the density of the former to decrease and deviate from the evolution observed
in the single-mass component model. However, the transition to this phase clearly occurs 
after the models have been already evolved for a time comparable (for $ f_{\rm bh}=5\times 10^{-3}$) or 
longer~(for $ f_{\rm bh}=10^{-3}$)  than the age of the Galactic NC \footnote{The
mean stellar age in the Galactic NC is estimated to be $\sim 5~$Gyr,~\citep[][]{figer+04}.}.

 Given the results of the simulations presented
in this section, we can schematically divide mass-segregation in two phases:
in phase (1) the density of BHs is smaller than the density of stars and the models  evolve mainly 
due to scattering off the stars -- the BHs inpiral to the center due to dynamical friction, and
the stellar distribution relaxes as in a single mass component model. 
In phase (2), when the density of BHs becomes  comparable and larger than the density of stars, BHs
and stars  evolve due to scattering off the BHs which causes the models to rapidly evolve toward the steady state.

Perhaps, the most  interesting aspect of our simulations is the long  timescale required by the BH population to 
segregate to center through dynamical friction (phase 1 above) and reach a (nearly) steady state distribution -- a time comparable
 to the relaxation time as defined by the dominant 
stellar population. In what follows we show that these predictions agree well with the evolution expected on 
the basis of theoretical arguments.

\begin{figure}
\centering
~~~~~~\includegraphics[width=0.35\textwidth,angle=0]{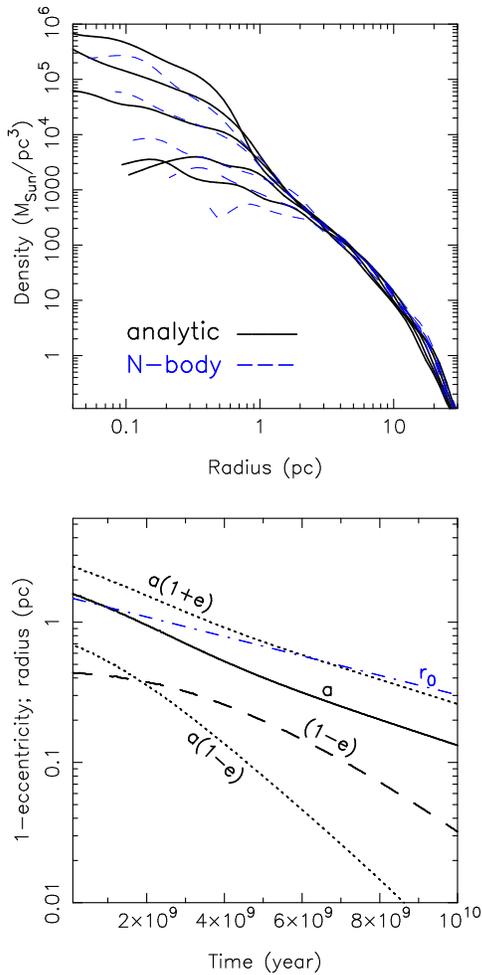} 
  \caption{The top panel shows the evolution of the density of $10~M_{\odot}$ BHs  due to dynamical 
  friction against the field stars. The density profiles are shown at $t=(0,2,4,6,8)~$Gyr. 
  Blue-dashed curves were obtained via direct $N$-body simulations; black-solid curves
  correspond to the results of our semi-analytical model in which the frictional force was computed 
  using Equation~(\ref{dfa3}), which accounts for the contribution of the fast-moving stars to the frictional
  force. Bottom panel shows the evolution of angular momentum ($1-e$) and semi-major axis ($a$) 
  of a $10M_\odot$ BH in our GC model. Blue-dot-dashed curve shows the break radius evolution
  of the background model.   Note that as the BH reaches roughly $r_0/2$ 
its orbital radius  migrates inward on a timescale similar to that
  over which the core in the background density evolves due to two-body relaxation.  }\label{mass-seg3}
\end{figure}

\subsection{Analytical estimates}
In order to understand the evolution of the distribution of BHs observed in the $N$-body simulations, we
evolved the population of  massive remnants using an analytical estimate for the dynamical 
friction coefficient. The stellar  background was represented as an  analytic potential which was also let evolve 
in time accordingly to  the evolution observed in the stellar distribution during  the $N$-body simulations. 

We began by generating random samples of positions and velocities from the isotropic distribution function corresponding to the density model of Equation~(\ref{den}). The orbital equations of motion were then integrated forward in time 
 in the evolving smooth stellar potential and including a term which describes the orbital energy dissipation due to dynamical 
 friction. 
 The dynamical friction acceleration
 was computed using the expression:
\begin{eqnarray} \label{dfa3}
\boldsymbol{a_{\rm fr} }  &\approx&  -4\pi G^2  m_{\rm bh} \rho(r) \frac{\boldsymbol{v}}{v^3}\nonumber \\
&&\times {\Big(} {\rm ln}  \Lambda \int_{0}^v dv_\star 4\pi f(v_\star) v_\star^2  ~
 \\
&& +\int^{{\infty}}_v dv_\star 4\pi f(v_\star) v_\star^2    \left[ {\rm ln} \left( \frac{v_\star+v}{v_\star-v} \right)-2\frac{v}{v_\star} \right]  \Big), \nonumber
\end{eqnarray}
with  $v$ the  velocity of the inspiraling BH and 
$f(v_\star)$  the velocity distribution of field stars.
 The second term in parenthesis of Equation~(\ref{dfa3}) represents 
the frictional force due to stars  moving faster than the test mass. 
Such ``non-dominant'' terms are neglected in the standard Fokker-Plank treatment in which the
dynamical friction coefficient is obtained by integrating only over field stars with velocity {\emph smaller}
 than that of the test particle~\citep[e.g.,][]{HA06,AH:09,M:10}.
\citet{AM12} showed that this approximation breaks down in a shallow density profile of stars around a MBH
where such terms can become dominant, as there are a few or even no particles moving more slowly than the local circular velocity. 

The $N$-body integrations show that the stellar distribution  changes with time  
in a quasi-self-similar way -- the stellar density profile break radius 
 shrinks progressively with time while the outer profile slope is maintained roughly unchanged. 
In order to account for such evolution,
 we computed at each time the best fitting density model of Equation~(\ref{den})
to the density profile of  stars in the $N$-body system at that time. 
We used this density model to compute
gravitational potential, distribution function and corresponding dynamical friction coefficient. 
This  procedure allowed us to include the evolution of the stellar background  when evolving  the BH population.
Our integrations are unique in the sense that they are the first including at the same time:
(i) a correct estimate of the dynamical friction coefficient, which takes into account
the contribution of stars moving faster than the inspiraling BH,  and (ii) a realistic treatment of the
evolution of the stellar background under the influence of gravitational encounters.
However, since our analysis does not take into account BH-BH self interactions,  our integrations are only valid 
until the density of BHs remains well below  the density of stars. In this respect, 
our approach is limited to the early evolution of the system, when the BHs only represent a negligible perturbation
on the evolution of the light component.

The upper panel of Figure~\ref{mass-seg3} shows the density profile of BHs obtained from  the semi-analytical modeling 
described above  and compares it to the results from the direct $N$-body simulation with BH fraction $f_{\rm bh}=10^{-3}$. 
The central density of BHs increases with time at a rate which is comparable in the  two models. 
 The plot shows that the spatial distribution of stellar-mass BHs near the GC 
might not have reached a steady state form -- at least if
their initial distribution was similar to what used in our models.
In fact, even after a time of $\approx 8$Gyr the central density of  BHs is still
 substantially lower than the density of stars.
As our analysis demonstrates, the persistence of such low densities of BHs  is a direct consequence of the long timescale
of inspiral in a density core near a MBH. This latter point is illustrated
in the lower panel of Figure~\ref{mass-seg3} which shows the trajectory  of a $10~M_{\odot}$ BH at the GC.
The rate of orbital decay  slows down as the BH reaches $\approx r_0/2$, due to the lack of low-velocity stars in the core.
After the BH reaches this radius dynamical friction becomes very inefficient and the decay of the BH orbit
proceeds at a rate which is comparable to the rate at which the core radius in the stellar distribution shrinks 
due to gravitational encounters --- a time of  order the relaxation time of the nucleus.

\subsection{Comparison with recent work}
In this section we used direct $N$-body integrations as well as analytic models to describe
 the evolution of multi-mass models of the Milky Way NC characterized by an initial parsec-scale core in the density distribution.
 Calculations similar to those described here were recently performed by \citet{preto-amaro10} and  \citet{GM12}.
  
\citet{preto-amaro10}  studied the evolution of models with two mass species: stars and BHs.
These authors concluded that mass segregation of the heavy component 
\emph{speeds up} cusp growth in the lighter component by factors up to 10 in comparison with the single-mass case.
This conclusion is somewhat in agreement with the results of our simulations which also show that a stellar cusp, extending out to roughly $r_{\rm infl}$, regrows faster in models with BHs~(e.g.,~Figure~\ref{mass-seg-slp}).
However, for realistic numbers of BHs we find that the timescale of cusp 
regrowth is only a factor of two shorter than in the single-mass component models.
\citet{preto-amaro10}   argued that the time scales associated with cusp regrowth are clearly 
 shorter than the Hubble time for nuclei similar to that 
of the Milky Way -- even though the relaxation time, as estimated for a single mass stellar distribution, exceeds 
one Hubble time.    Based on our study we conclude instead that
over one Hubble time and if a standard IMF is adopted adding a heavy component has relatively little effect on the evolution 
of the main-sequence component (e.g., Figure~\ref{mass-seg2}). 
Even for a top-heavy IMF, which results in initial larger densities of BHs,  
the time for cusp regrowth is longer than the mean stellar age in the Galactic center ($\sim5~$Gyr). 
The reason for this is that 
due to the inefficient dynamical friction force in a density core around a MBH,  the central density of BHs 
remains well below the density of stars for a time of order the relaxation time of the nucleus.
The time required to regrow a cusp in the stellar distribution appears to be longer than the Hubble time for galaxies
 similar to the Milky Way.

\citet{GM12} simulated the merger between galaxies with MBHs containing four mass groups, representative of old stellar
populations. They followed  the evolution of the merger products for about three relaxation times and
 found that the density cores formed during the galaxy mergers persisted, 
and that the distribution of the stellar-mass black holes evolved 
``against an essentially fixed stellar background". \citet{GM12}  
also integrated the exact same Fokker-Planck models as in  
\citet{preto-amaro10} and argued that the accelerated cusp growth described by 
these latter authors  is seen to be present only at small radii, $r\lesssim 0.05~{r_{\rm  infl}}$.
 At radii larger than these adding
the BHs has the effect of  lowering the density of the stellar component at all times.
 \citet{GM12}  argued that \citet{preto-amaro10} were misled by looking at the very-small-radius 
 regime in their Fokker-Plank solutions, where the cusp in the main sequence component stands out. 
  Our study shows  that the BH population has indeed two effects on the main-sequence population: 
 it lowers the ``mean'' density of stars  \citep[the point stressed in][]{GM12}, and 
 it accelerates the redistribution of the stars in phase-space, toward the $\gamma=3/2$ steady-state
 \citep[as found in][]{preto-amaro10}. So in 
 a sense, the BHs both ``create" and ``destroy" a cusp: although 
 the presence of a BH population can significantly accelerate the timescale of cusp 
 regrowth in the stellar distribution, the scattering off the heavy
(BH) component causes the density of stars to decrease at radii larger
 than $\sim 0.05~{r_{\rm  infl}}$.

 \section{Globular cluster merger model; evolution Time Scales}\label{timescales}
In the previous section we have shown that due to the long timescales of evolution,  the current distribution of BHs and stars 
at the center of galaxies similar to the Milky Way should be considered very uncertain. 
  In these and more massive galaxies the current distribution of stars and BHs can still reflect  their initial conditions
 and the processes that have lead to the formation and evolution of their central NC.
 This conclusion suggests that standard mass-segregation models, 
 which assume the same initial phase space distribution
for BHs and stars, can lead to misleading  conclusions regarding the current dynamical state of galactic nuclei and
motivates studies that start from initial conditions which correspond to  well-defined physical models.

 In what follows, we  present a set of  $N$-body experiments which were designed to
understand  the distribution of  stars and BHs in galactic nuclei 
formed via  repeated merger of massive stellar clusters -- 
a formation model which has been shown to be very successful in reproducing the observed properties and scaling
relations of nuclear star clusters~\citep[e.g.,][]{turner+12,me13}.
We begin here with discussing the relevant timescales 
of the problem, including the characteristic orbital decay  time of massive clusters in the inner regions of galaxies,
and   the relaxation timescales of  galactic nuclei and globular clusters.
  
\subsection{Globular Clusters decay time.} 
\begin{figure}\centering
 \includegraphics[width=0.4\textwidth,angle=0]{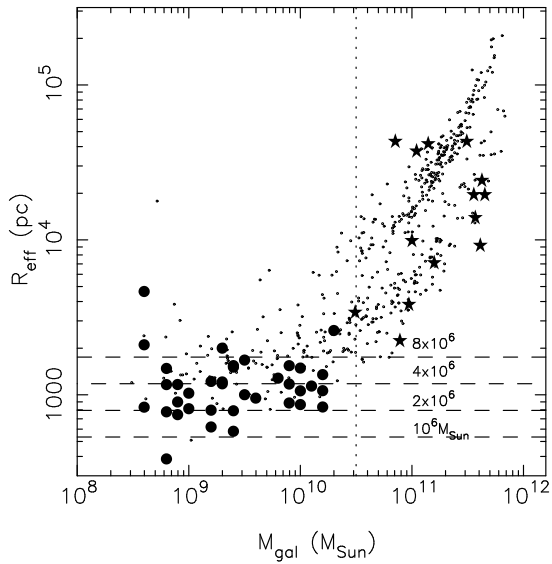}
\caption{effective radii~($R_{eff}$) of galaxies plotted against their
masses~($M_{gal}$). Filled circles and star symbols represent respectively nucleated and un-nucleated
galaxies that belong to the Virgo galaxy cluster~\citep{Cote}. 
Dots are data from~\citet{FORBES}.
Vertical line gives the value of the mass ($M_{gal} \approx 10^{10.5}~M_{\odot}$) which 
approximately set the transition between nucleated galaxies ($M_{gal} < 10^{10.5}~M_{\odot}$),
and MBH dominated galaxies~($M_{gal} > 10^{10.5}~M_{\odot}$)~\citep{Cote}.
Dashed-horizontal curves give  the value of $R_{eff}$ obtained by setting $\Delta t_{h}=10^{10}~$ in  Equation~(\ref{df})
and adopting various masses of the sinking objects. The large effective radii of massive galaxies 
give time scales for inspiral that are usually longer than one Hubble time.
 The observed lack of nuclei in massive galaxies 
could be  explained as a consequence of  the  longer infall times in these galaxies, due to their 
large effective radii.
 }\label{rvsm-gxs}
\end{figure}

Sufficiently  massive and compact clusters can decay towards their parent
galaxy central region in a time much shorter than the Hubble time.
An approximation  of the time for clusters (within the half mass radius of the stellar bulge)
to spiral to the center is given by~\citep{AM12-2}:
\begin{equation}\label{df}
\Delta t_{h}  \approx 3  \times 10^{10} {\rm yr}~~ \tilde{R}_{eff}^{1.75}  \tilde{m}^{-1}~.
\end{equation}
with  $\tilde{R}_{eff}$ the galactic effective radius in kpc, and $ \tilde{m}$  the globular cluster  mass in units of
$10^6~M_{\odot}$.   Within $\Delta t_{h}$ the forming nucleus has a luminosity comparable to that of the surviving clusters.
Equation~(\ref{df})  predicts that a significant fraction of the globular cluster population in  faint 
and intermediate luminosity stellar spheroids would have spiraled to the center by now,
while in giant ellipticals, due to their larger characteristic radii, the  time required to grow a NC might be longer than $10^{10}~$yr.
We stress that the inspiral time obtained using  Equation~(\ref{df})  gives only a crude approximation (likely an overestimate) of the real 
dynamical friction time scale. 
Nevertheless it is reasonable to draw the conclusion
that the observed lack of nuclei in galaxies more massive than  about  $10^{10}~M_{\odot}$
could be due to the  longer infall times in these galaxies, due to their 
larger values of $R_{eff}$. 
Figure~\ref{rvsm-gxs} 
presents a test of this idea. This figure gives effective radii  versus masses, $M_{gal}$, for galaxies 
belonging to the Virgo cluster that either have~(filled black circles) or do not have~(star symbols) a central NC. Dashed curves 
give the value of $R_{eff}$ obtained by setting $\Delta t_{h}=10^{10}~$yr in  Equation~(\ref{df})
and adopting various masses of the sinking object.  The figure shows that
  only in galaxies with $M_{gal}\lesssim 10^{10}~M_{\odot}$,  massive globular clusters would have
 enough time to spiral into the center, merge and form a compact nucleus. 
The observed absence
 of compact nuclei in giant ellipticals could be therefore interpreted as a consequence 
 of the long dynamical friction time scale of globular clusters in these galaxies.
 
 We add that the density profile of stars in giant ellipticals  is often observed to be flat or 
slowly rising inside the influence radius of the MBH. As shown in Section~\ref{sec-mass-seg} this  implies a very long dynamical friction
time scale  inside the MBH influence radius due to the absence of stars 
moving more slowly than the local circular velocity~\citep{AM12}.
Massive   clusters orbiting within the core of a giant elliptical galaxy  do not reach the 
  center  even after $10^{10}~$yr. In addition, due to the strong tidal field produced by the MBH, 
 globular clusters can only transport little mass to the very central region of the galaxy.
  Both these effects, i.e. long inspiral times and little mass transported to the center,
have been suggested to suppress  the formation of  NCs in bright galaxies, in agreement with observations~\citep{me13}. 
 
%

\subsection{Nuclear star clusters relaxation time}
A useful reference time for our study  is the relaxation time computed at the radius containing half of the mass of the system, $R$. Setting
${\rm ln} \Lambda=12$,  and ignoring the possible presence of a MBH, the half mass relaxation time is:
\begin{equation}\label{th}
T_h=
2.1 \times 10^5\frac{ \left[ r_h(\rm pc)\right]^{3/2} N^{1/2}}{(m/M_{\odot})^{1/2}}\frac{10}{\ln (0.4 N)}{\rm yr}~,
 \end{equation}
 where $N$ is the total number of stars.

In the absence of a MBH, collisional relaxation leads to mass segregation and core collapse.
 In a pre-existing NC, the presence of a MBH inhibits core collapse, causing instead
the formation of a Bahcall-Wolf cusp,~$n \sim r^{-7/4}$, on the two-body relaxation time scale \citep{PR04,M09}. 
 Nuclear clusters  belonging to the Virgo galaxy cluster have half-mass relaxation time that scales with the total 
 absolute magnitude of the host galaxy, $M_B$, as~\citep{M09}:
 \begin{equation}\label{hmrt}
{\rm log}\left( T_h/{\rm yr}\right)=9.38-0.43(M_B+16)~.
 \end{equation}
  Galaxies with luminosities less than $\sim 4 \times 10^8~L_{\odot}$, have NCs with relaxation times that fall below 
 $10~$Gyr. These galaxies have NCs with masses $\lesssim 10^7~M_{\odot}$ and half mass radii $r_h\lesssim 10~$pc.
 These limiting values appear close to those characterizing the Milky Way NC, suggesting that only spheroids fainter 
 than the Milky Way have collisionally relaxed nuclei.     Relaxation times for nuclei with masses $\gtrsim10^7~M_{\odot}$
 are therefore too long for assuming that they have reached a collisionally relaxed state, but 
 they are still short enough that gravitational encounters would substantially affect their
structure over the Hubble time. This is in agreement with the results of Section~\ref{sec-mass-seg} and also appears
to be consistent with absence of a Bahcall-Wolf cusp in the distribution of stars at the GC.
 


\begin{figure}
 \includegraphics[width=0.24\textwidth,angle=270]{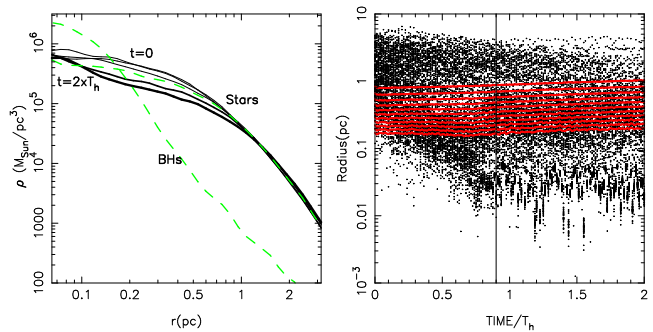}
\caption{Evolution of the  cluster model during the initial $N$-body integrations 
used  to realize the mass-segregated cluster models for Runs~A1 and A2~(see Table~1).
The left panel shows the evolution of the stellar density profile. Line thickness increases with time. The green-dashed curves 
give the density profile of stars and BHs used as initial conditions for the inspiral runs.
Right panel gives the Lagrangian radii of the stellar component~(red solid curves) and the 
BHs distance from  the cluster center~(black dots). The BHs segregate to the cluster core
forming a dense sub-cluster in about one, half-mass relaxation time as defined by 
the stellar component. The stellar cluster slightly expands due to heating by the inspiraling BHs.
Vertical line gives  the time at which we extracted the initial conditions for the inspiral simulations.
}\label{mscluster}
\end{figure}


\subsection{Globular clusters relaxation time}
Globular clusters with $N\sim 10^{6-7}$ have relaxation times $T_h\sim10^{9-10}~$yr. 
Most Galactic globular clusters are therefore relaxed systems.
The  time scale required for the BH population to segregate to cluster center and there form a subcluster 
dynamically decoupled from the host stellar cluster is
approximately the core collapse time for the initial BH cluster~\citep[e.g.,][]{BBK10}:
\begin{equation}\label{th2}
T_{ms}\approx \frac{m}{m_{bh}}T_{cc}~,
 \end{equation}
where ${m_{bh}}$ is the mass of a stellar black hole and $T_{cc}$ is the core collapse time of the host stellar cluster
which is about $T_{cc}\approx 15\times T_{h}$ for a Plummer model.
After  $\sim T_{ms}$ the central density of BHs becomes large enough that BH-BH binary formation 
takes place through three or four body interactions~\citep{HH03}.  
The formed BH binaries then ``harden" through repeated super-elastic encounters that  lead to the ejection of BHs
from the cluster core until eventually only a few BHs are left~\citep[e.g.,][]{BBK10}.

In galaxies similar tot he Milky Way, stellar clusters with masses $\gtrsim 10^6~M_{\odot}$ and starting from a 
galactocentric radius of $1~$kpc  have orbital decay times due to dynamical
friction less than $\lesssim 3~$Gyr~(Equation\ref{df}). 
The clusters dynamical friction time is therefore typically long compared to 
the timescale over which the BHs would segregate to center of the cluster. 

It is possible that the BH population will evaporate  through super elastic encounters
before the cluster reaches the center of the galaxy. This could lead to the formation 
of a NC with a much smaller abundance of BHs relative to stars than what predicted by standard initial mass functions.
On the other hand, for massive clusters after the BHs are already segregated to the center,  the 
encounter driven evaporation time scale of the BH sub-cluster typically requires an additional 
few Gyr of evolution to complete  ~\citep{DBGS10,DBGS11}. 
 Moreover, recent theoretical studies~\citep{morscher+13,sippe+hurley13}, together with several  observational 
evidences~\citep{maccarone+07,brassington+10,maccarone+11,strader+12},  show that
  old globular clusters may still contain hundreds of stellar BHs at present which suggests that  BH depletion 
  might  not be  as efficient as previously thought. This
  indicates that for many large clusters (the ones most relevant to NC formation),
   most of the BHs will  not  be ejected   
   before inspiral has occurred.
The above  arguments convinced us  that  the inspiral of massive clusters 
in the central region of the Galaxy could serve as a continuos source term  of 
BHs in these regions.

 \begin{table}
 \begin{center}
\caption{Initial Modles \label{t1}} 
\begin{tabular}{llllll}
 \hline
 ${\rm Run}$  & $N_{\rm gal}$ &  $N_{\rm cl,\star}$ &  $N_{\rm cl,bh} $ & \# of infalls & Galaxy model \\ 
 \hline 
 A1      &      $1     \times  10^6$ &  45720  & 240     &  1  &	 Model~1	\\
 A2      &      $1.5  \times  10^6$ &  45720  & 240     &  1	&   Model~2      \\
 B        &      $4.6  \times  10^5$ &  5715     & 33       &  12	&   Model~1	 \\	
 C        &      $4.6  \times  10^5$ &  5715     & 100    &  12 	&   Model~1	 \\
 \hline
 \\ \\
\end{tabular}
\end{center}
\end{table}

\section{Numerical set-up}\label{ics}
\subsection{Initial conditions and numerical method}
In \citet{AM12-2} we used $N$-body simulations to study how the presence of a MBH at the center
of the Milky Way impacts the globular cluster merger hypothesis for the formation of its NC. We determined 
the properties of the stellar distribution in a galactic nucleus forming through 
the infall and merging of globular clusters. We showed that  a model in which a large fraction 
of the mass of the Milky Way NC arose from infalling  
globular clusters is consistent with existing observational constraints.  
Here we replaced the single-mass globular cluster models
of \citet{AM12-2} with systems containing (in addition to the stellar component) a  remnant population of BHs.

These simulations  were performed by using $\phi$GRAPE \citep{HGMM0}, 
a direct-summation code optimized for running on GRAPE 
accelerators \citep{MakinoGRAPE}. This integrator is equivalent  
to $\phi$GRAPEch which we used in Section~\ref{sec-mass-seg}, but without the regularized chain.
The accuracy and performance of the code are  set by the time-step parameter $\eta$ and
the softening  length $\epsilon$.  
We used a Plummer softening for the gravitational force between  particles, 
and we did not model binary formation in the calculation reported below. 
We set $\eta=0.01$ and $\epsilon=10^{-4}$pc.
With these choices, energy conservation was typically
$\lap  0.01\%$ during each merging event.
The simulations were  carried out using  the  32-node  GRAPE cluster at the Rochester Institute of Technology,
and also on Tesla C2050 graphics processing units on the 
Sunnyvale cluster at the Canadian Institute for Theoretical Astrophysics.
In the latter integrations, $\phi$GRAPE was used in serial mode with {\sc sapporo}~\citep{sap}.
Each simulation required  between $3$ to $4$ months total of computational time.

Table~1 summarizes the parameters of the $N-$body models. We performed  four simulations. 
Runs A1 and A2 are high resolution simulations ($N\sim 10^6$) that explore 
 the dynamics of one single globular cluster inspiral. In simulations B and C the total number of $N$-body particles
was greatly reduced  in order to more efficiently  follow the consecutive inspiral and merger of 
$12$ dynamically evolved  clusters.  In these latter simulations each inspiral simulation
was started  after the stars from the previously disrupted cluster  were 
set to a state of collionless dynamical equilibrium and the number of BHs in the cluster remnant dropped to 
$<10$. This corresponds to  a time of   $1-3\times10^7$yr
for each inspiral event to complete,  with the longer times corresponding to the earlier infalling clusters.
The clusters were  initially  placed on  circular orbits at a distance of $20$~pc from the center.
The choice of circular orbits was motivated by  the well-known effect of orbital
circularization due to dynamical friction \citep[e.g.,][]{CPV,Hashimoto}.
In the consecutive merger simulations~(runs~B and C), in order not to favor any particular direction for the inspiral, the orbital
angular momenta were selected in the following way:
the surface of a sphere can be tessellated by means of 12
regular pentagons, the centers of which form a regular dodecahedron
inscribed in the sphere.
The coordinates of the centers of these pentagons were identified with the
tips of the 12 orbital angular momentum vectors.
In this way, the inclination  and longitude of ascending node  of each initial
orbit were determined.

 \subsection{Galaxy models}
We adopted two different $N-$body models to represent the central region of the Galaxy. 
Model 1 is obtained by an inner extrapolation
of the observed density profile of stars in the Galactic nuclear bulge outside $10~$pc.
In these regions the Galaxy is dominated by the presence of the nuclear stellar disk
which  is characterized by a flat density profile.
Accordingly, we  adopted the truncated shallow power-law density model:
\begin{equation}\label{dm}
\rho_{\rm gx,1}(r)=\tilde{\rho}\left(\frac{r}{\tilde{r}}\right)^{-\gamma}\text{sech}\left(\frac{r}{r_{\rm cut}}\right)
\end{equation}
where $\tilde{\rho}=400$M$_{\odot}/$pc$^{3}$  is the density at  $\tilde{r}=10$pc, and the truncation radius is 
$r_{\rm cut}=20~$pc.
Hence, the initial conditions for  Model~1 does not include a preexisting NC and they correspond to a shallow
density cusp  around a central MBH.

Model~2 was obtained  by the superposition of the density model of Equation~(\ref{dm})
and a broken power law  model representing a NC:
\begin{eqnarray}\label{brokenpl}
\rho_{\rm gx,2}(r)&=&\rho_{\rm gx,1}(r) \nonumber \\ 
 &+& \rho_b\left(\frac{r}{r_{b}}\right)^{-\gamma_i} 
\left[1+\left(\frac{r}{r_{b}}\right)^\alpha\right]^{(\gamma_i-\beta)/\alpha}\text{sech}\left(\frac{r}{r_{\rm cut}}\right),~~~~
\end{eqnarray}
with $\rho_b=4.1\times10^4\text{M}_\odot/\text{pc}^3$, $r_b=1.5~$pc,~$\gamma_i=0.5$, $\beta=1.9$,  $\alpha=3.73$ and $r_{\rm cut}=20~$pc.
This model corresponds approximately to the best fitting density profile  of  the simulations end-product   of~\citet{AM12-2}.

In both Model~1 and Model~2 we included a central  MBH of mass~$M_{\bullet}=4\times 10^6~M_{\odot}$ and we 
generated their $N$-body representations via numerically calculated distribution functions.
  
\subsection{Star clusters model}
We generated our globular cluster initial conditions following
the same procedure described in \citet{AM12-2}, where a detailed description of the initial conditions of
the clusters  can be
found.  In brief, the clusters are started on circular orbits of radius $r=20~$pc, and their initial masses and radii are set up in such a way as to be consistent with the galactic tidal field at that radius.
The clusters are King models~\citep{K62} with central (King) potential $W_0$=5.8,
core radius $r_k$=0.5~pc, and central velocity dispersion $\sigma_K=35~{\rm km~s^{-1}}$.
With this set of parameters the truncated mass of the clusters was
$m_{\rm cl}\approx 1.1\times10^6~M_{\odot}$.

To these models  we added   a heavier mass group representing a population of stellar BHs.
The relative values of the particle masses was 1:10. These represent respectively one solar mass stars and $10~M_{\odot}$ BHs. The two mass groups had the same initial phase-space distribution.
Standard population synthesis models  predict that about $1\%$ of the total mass in a stellar system will be in  BHs, 
top heavy mass functions result in about  five times more BHs.
Accordingly,  in our initial models the total mass in BHs was $10^{-2}$  and $5\times10^{-2}$ times 
$4\times 10^6M_{\odot}$ (the
mass of the \emph{ non-truncated} King model) for run A-B and C~(Table~1) respectively. 
The choice of scaling the total number of BHs to the initial non-truncated cluster mass is based on the fact that
when the cluster reaches $100~$pc~(roughly the radius at which  our models start to be tidally truncated) 
mass segregation is likely to have already occurred. under these circumstances, tidal stripping will preferentially remove stars from the outer part of the system,
leading to an overabundance of BHs with respect to standard mass functions~\citep{BK11}.
In addition, this choice resulted in a sufficiently good statistics for the remnants population.
The effect of varying the initial mass in stellar  BHs and their initial dynamical state inside the clusters will be  investigated in a future paper.

The mass-segregated cluster models were then created via $N$-body integrations, starting from 
 the cluster equilibrium models. 
 Figure~\ref{mscluster} gives the evolution of  the two mass components during these integrations.
We let the system evolve for a few relaxation times as defined by Equation~(\ref{hmrt}).
The stellar BHs accumulate toward the center and by approximately one 
half-mass relaxation times their distribution appears to have reached an approximately steady state. 
At the same time  the density profile 
defined by the stellar component undergoes a slow expansion  due to heating by the BHs.
 
  \section{Results}\label{results}

\begin{figure*}
\centering
	\includegraphics[width=.964\textwidth]{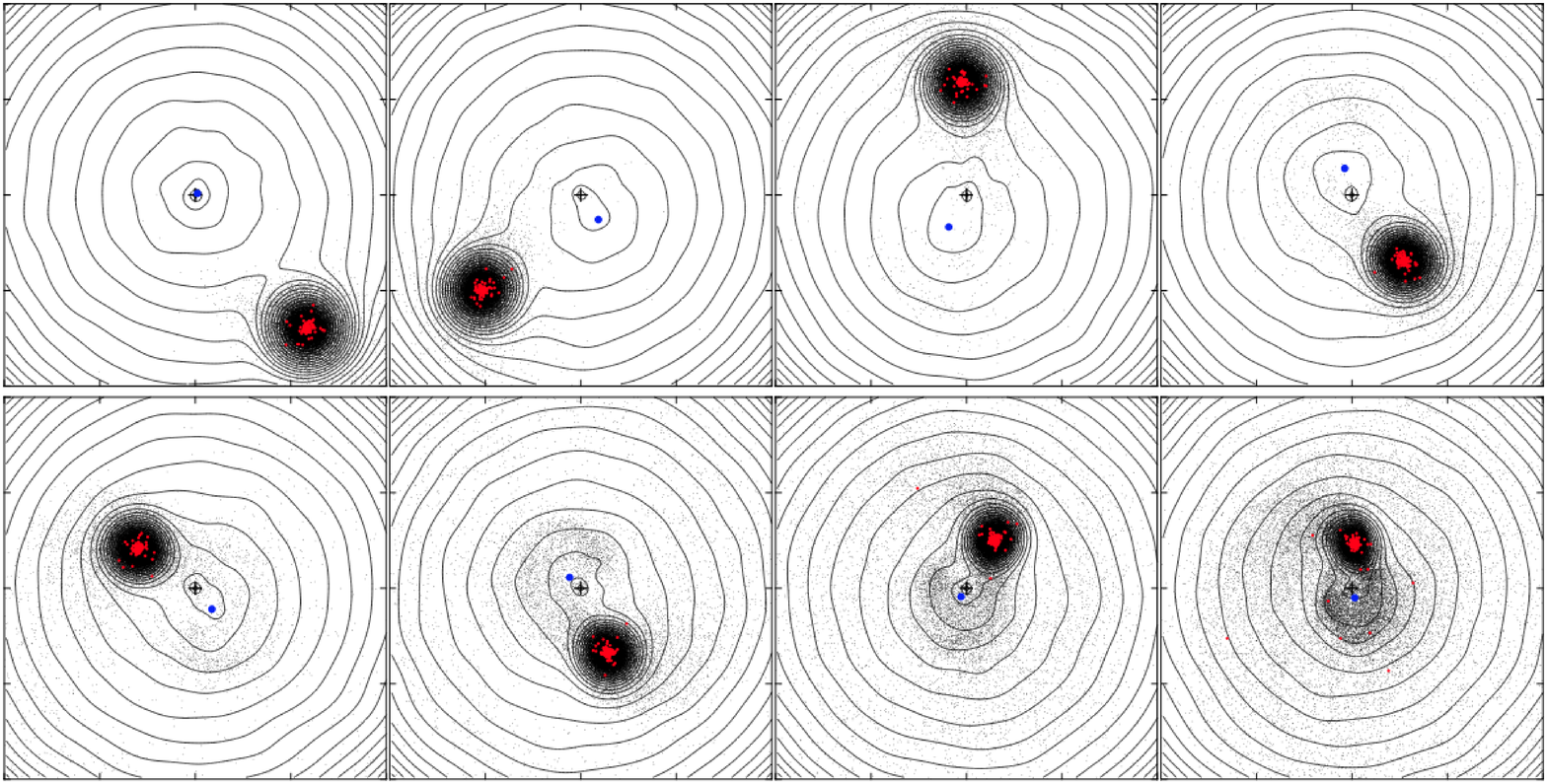}\\
	~\includegraphics[width=.971\textwidth]{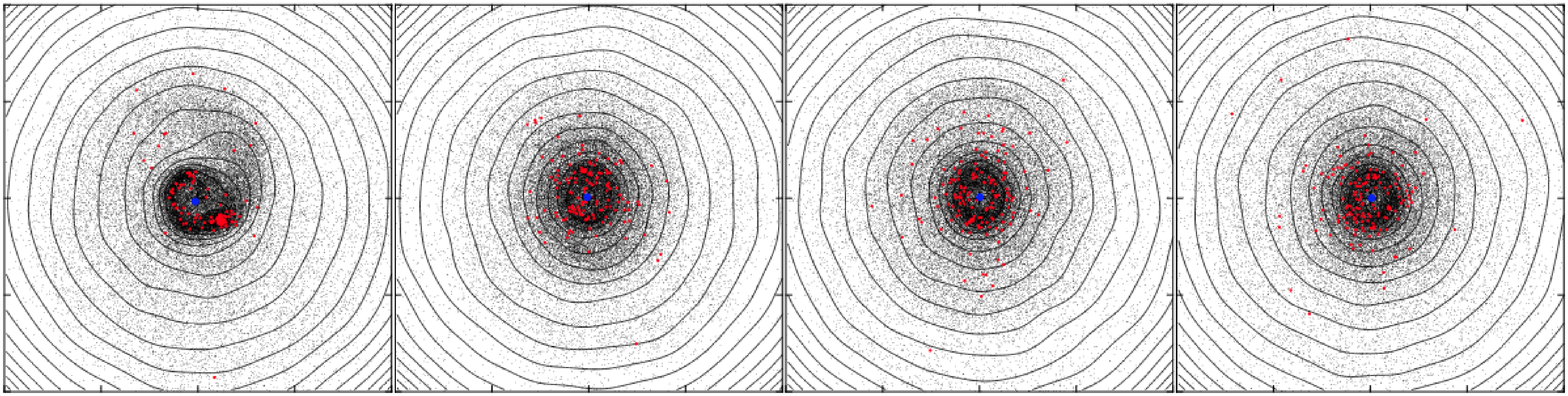}   
\caption{Inspiral of a $\sim10^6~M_{\odot}$ globular cluster in a Milky Way model. 
The linear size of each box is $20~$pc; the time separation between each snapshot
is $5\times 10^5~$yr. The blue filled circle marks  the galactic MBH position.
Curves show the contours of the projected density of the background galaxy.
 Red points represent the  globular cluster BHs; black dots the cluster stars.}\label{f2snap}
\end{figure*}

\begin{figure}
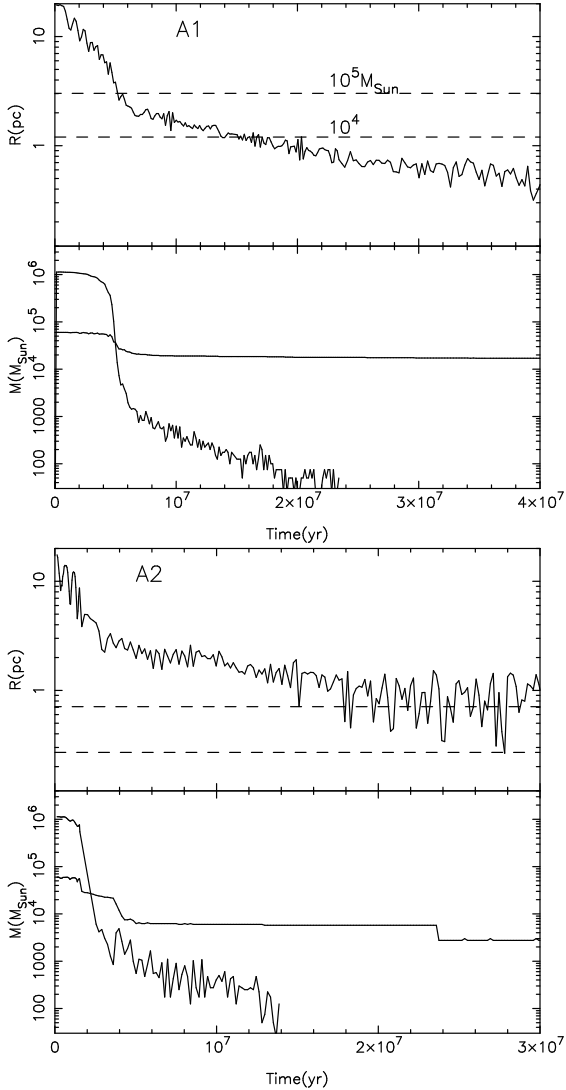

\centering
\includegraphics[width=.4\textwidth,angle=270]{Figure8a.ps}
\includegraphics[width=.4\textwidth,angle=270]{Figure8b.ps}
\caption{Time evolution of  galactocentric radius~(upper panels)
and bound mass~(lower panels) of the globular cluster models in
runs A1 and A2~(see Table~\ref{t1}). 
In the lower panels we show separately  the bound mass in BHs (curve starting at $6\times10^4M_{\odot}$) and 
in stars~(curve starting at $1.1\times10^6M_{\odot}$)
Dashed curves in the upper 
panels give the galactocentric radii in the initial galaxy models containing $10^4$ and 
$10^5~M_\odot$.}\label{Fig-mvst}
\end{figure}

 \begin{figure}
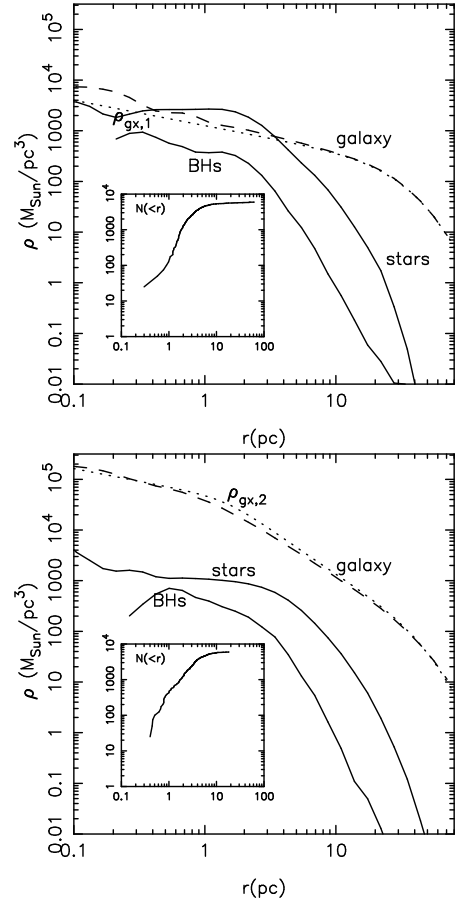

\centering
 \includegraphics[width=0.33\textwidth,angle=270]{Figure9a.ps} \includegraphics[width=0.33\textwidth,angle=270]{Figure9b.ps} \\
\caption{Density profile of BHs and stars at the end of the single inspiral simulations of Table~1. 
Upper panel corresponds to simulation A1 and lower panel to simulation A2.
Density profiles are 
given after the BH cluster are artificially dispersed  as described in the text.
Insert panels show the corresponding cumulative number distributions of stellar BHs. Dotted curves give
the density model of the initial galaxy models of Equation~(\ref{dm}) and Equation~(\ref{brokenpl}). The density 
profile of the galaxy at the end of the simulation is shown as dashed curves.
Up to $\sim 100$ BHs are migrated inside the inner $0.1-0.2~$pc of the galactic center. 
}\label{density-single-inpiral}
\end{figure}

 \subsection{Single inspiral simulations}\label{single}
 Figure \ref{f2snap} shows  surface density contours of  the single inspiral simulation A1.
After $\sim 10^7~$yr the stellar cluster is  at about  $5~$pc from the center; 
at these distances the  disruption process due to 
 tidal stress from the central MBH begins. 
 Rapid removal  of stars from the outer part of the cluster by the galaxy and MBH tidal fields unveils its 
 mass-segregated  BH cluster.
Figure~\ref{Fig-mvst} shows the time evolution of radius and 
bound mass for the globular clusters in  runs A1 and A2.  
Our cluster models rapidly evolve to a state of dark stellar cluster, i.e., 
a dense cluster dominated by dark stellar remnants~\citep{BK11}.
After this state is reached, due to the drop in the total cluster mass~(see lower panels of Figure~\ref{Fig-mvst}), 
the dynamical friction drag on the remaining BH cluster  is largely suppressed, slowing down 
its orbital decay toward the center of the galaxy. 
Noticeable, due to the common  motion around the  system MBH-cluster center of mass, the MBH is significantly 
 displaced from the galaxy center. More precisely, we found a maximum  displacement of $\sim5~$pc  in run A1
and a somewhat smaller (maximum) displacement of $\sim2~$pc in run A2.

Figure~\ref{Fig-mvst} shows that after the stellar clusters are disrupted, the remaining dark clusters  
 have a bound mass of $\sim 2.7\times 10^3~M_{\odot}$  in simulation A1 and 
$1.7\times 10^4~M_{\odot}$ in simulation A2, corresponding to  9 and 67 BH particles respectively.
The enhanced removal of stars and BHs decelerates the orbital evolution of the cluster due to its lower mass.
During the inspiral after about $2\times 10^7~$yr the  BH cluster core  has
collapsed to $\lesssim 0.05~$pc. This makes the central density much higher, which prevents the complete disruption of the BH cluster. 

Assume that the  cluster  has reached a state of ``thermal equilibrium'' at the center, 
i.e., a state in which the stars and BHs are represented by lowered
Maxwellians: $m_{\rm st}\sigma_{\rm st}^2=m_{\rm bh}\sigma_{\rm bh}^2$, with
 $\sigma_{\rm st}$  ($\sigma_{\rm bh}$) the central one-dimensional  velocity dispersion
of the stars (BHs). If a MBH of mass $M_{\bullet}$ is present at the center of the galaxy, 
disruption occurs at  distance 
\begin{equation}
r_\mathrm{disr} \approx 2
\left(\frac{\sigma_\mathrm{NC}}{5\sigma_K}\right)^{2/3}
\left(\frac{r_\mathrm{infl}}{r_K}\right)^{1/3} r_K~,
\end{equation}
from the MBH. 
Then the BH cluster tidal disruption radius will be  smaller than that of the stellar cluster 
 by a factor
\begin{equation}
\approx 0.3 \left( \frac{1}{20}  \frac{r_{K; \rm bh}}{r_{K; \rm st}}  \right)^{2/3}  \left(  \frac{1}{10} \frac{m_{\rm bh}}{m_{\rm st}} \right)^{1/3},
\end{equation}
which suggests that the BHs can end up being more centrally concentrated than the stars.
A  condition for this to happen is that the BH cluster must not evaporate before
it has lost significant orbital energy by dynamical friction. 
In fact, the internal evolution of a compact  BH cluster  embedded in the extreme tidal field of the GC
 can proceed very rapidly and lead to the cluster complete dissolution  
on a short timescale, of order a few Myr~\citep[e.g.,][]{BK11,gurkan+05}.
 In our simulations the internal dynamical evolution of the clusters
 has been suppressed by giving a non-negligible softening radius  to the cluster particles.
In fact, the adopted integrator  cannot treat the postcollapse evolution of the cluster, since we used a softened potential.
 Thus we terminated the simulation at $\approx 3\times10^7$yr of evolution after the clusters have been fully depleted of stars and the remaining BH clusters underwent core-collapse.

 The end-product spatial density profile and cumulative mass distribution
of stars and BHs are  given in Figure~\ref{density-single-inpiral}. 
In order to obtain the BH density profile we forced the unbinding of the remnant clusters after $10^7$yr of evolution.
The unbinding of the clusters was induced
by ``turning off'' the gravitational interaction terms between the BH cluster members and by letting the system evolve for
about one crossing time. Although quite artificial, this procedure allowed us to account for the
 fact that the dissolution time of the cluster remnants is expected to be short relative to  
 their dynamical friction timescale -- in our models a $10^4 M_\odot$  system starting from a galactocentric radius of $1$pc reaches a 
radius of $0.2~$pc after $10^8$yr. The BH clusters will dissolve on a timescale proportional to the half-mass relaxation time,
 $T_{ev}\approx 300\times T_h$~\citep{Spitzer} (if we ignore the effect of the external tidal field); 
 from Equation~(\ref{th}), using $r_h=0.1$pc~(Figure~\ref{mscluster}),
$N=1000$, and $m=10M_{\odot}$ we find $T_{ev} \approx  10^7$yr. 
We stress that since the current state of the art computational capability does not allow us to simulate the 
actual number and mass of stars and to calculate the internal evolution of the cluster,  the densities of BHs
obtained here should be only considered as approximate (likely an underestimate of the real density).
Note also that it is unlikely that the core collapse phase of a BH cluster can lead to the formation of an
intermediate mass black hole since any BH-BH merger will eject the remnant from the cluster via asymmetric emission
of gravitational wave radiation before it can accrete other BHs or sourradning stars.

The number of BHs in our simulations, $N_{\rm BH, nb}(<r)$, was converted to a predicted 
number for a Milky Way model, using the approximate scaling:
\begin{equation} \label{bh-sc}
N_{\rm bh}(<r)=N_{\rm BH, nb}(<r) \times  \frac{4\times 10^6 M_{\odot}}{10 M_{\odot}}  \frac{m_{\rm BH, nb}}{M_\bullet}
\end{equation}
with the last factor at second term containing the masses in the units of the $N$-body code.
The number of BHs transported  in the inner $\sim1~$pc is  of order $100~$  in both A1 and A2. 
Our simulations result in a mass distribution characterized by a 
flat density core inside $\sim2$pc in run A1 and $\sim3$pc in run A2, and an envelope the falls off rapidly at large radii.
The BH density distributions  flatten within a radius 
comparable to the size of the core observed in the stellar density profile. This is because the BH cluster 
does not experience significant orbital decay after the star cluster is fully dispersed.
The difference in the mass distribution in the two models A1 and A2
is caused by the difference in the enclosed mass in the background galaxy. 
The tidal field near the galactic center is much stronger
in simulation A2 due to presence of a pre-existing compact stellar nucleus. 
 The stronger tidal field in the galaxy model of simulation A2 
 results in a larger core in the density distribution due to the larger tidal disruption
 radius of  the star cluster.

The dashed curves in  Figure~\ref{density-single-inpiral} 
display the density profile of the galaxy background at the end of the simulations.
A comparison  with the functional forms of Equation (\ref{dm}) and (\ref{brokenpl}) used to generate the initial
equilibrium models shows that the background galaxy  in A1 did not  evolve  appreciably;
in A2, instead, the final density of the galaxy appears to have slightly changed showing higher central densities and 
a smaller core radius ($\sim0.5~$pc) than  the initial model.
Thus the influence of the  inspirals on  the pre-exisiting distribution of field stars in our models is negligible at large radii $\gtrsim 3$pc, while leads to slightly higher central densities within this radius and a smaller core  radius relative to the initial model distribution~(we discuss this point in more detail below in Section~\ref{acc-rel}).

 \begin{figure}
\centering
 \includegraphics[width=0.33\textwidth,angle=0]{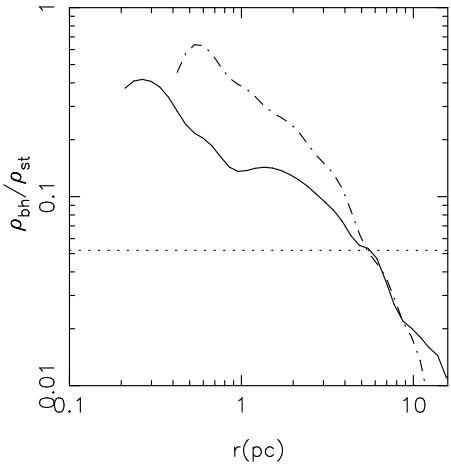}
\caption{Radial profile of the ratio  between the density of BHs, $\rho_{\rm bh}$,
and stellar densities,  $\rho_{\rm st}$, for simulations A1~(continue curve) and A2 (dot-dashed curve). The dotted curve
shows  the value corresponding to the initial cluster models before the BHs segregated to the center.
This latter is also the density ratio  expected in the absence of dynamical evolution. 
Within $5~$pc our models predict larger BH densities than
what expected if the remnant population had the same density distribution of stars at the 
time the clusters reach their tidal disruption radius. As discussed in the text, 
where $\rho_{\rm bh}\gtrsim 0.1\rho_{\rm st}$
the evolution of the stars is expected to be dominated by 
gravitational interaction with the BHs -- as oppose to self-interactions.
  }\label{density-cont}
\end{figure}

The key question to be answered by these simulations is the degree to which 
the density of stellar BHs near the center of the galaxy is enhanced, after the inspiral, 
with respect to the relative density expected in the absence of dynamical evolution.
Figure~\ref{density-cont} shows the radial profile of the ratio, $\rho_{\rm bh}/\rho_{\rm st}$,
of BHs to stellar densities. The dotted curve in the figure gives the density ratio, 
$\rho_{\rm bh}/\rho_{\rm st}=0.052$, of the initial  model before the BHs segregate
to the cluster center, which is also  the relative BH density expected in the absence of the cluster  dynamical evolution. 
Inside $5$pc our simulations result in larger BH densities relative to what expected if
the remnant population had the same density distribution of stars at the moment 
the clusters reach their tidal disruption radius.

Figure~\ref{density-cont} suggests that the evolution of a NC formed through the
merger of dynamically evolved  stellar clusters will be dominated by the BHs.
 The condition that the evolution of the light component 
 is dominated by  scattering off the heavy component
is \citep[e.g.,][]{GM12}:
\begin{equation}\label{bh-dom}
\rho_{\rm bh}\gtrsim (m_{\rm st}/m_{\rm bh}) \rho_{\rm st}=0.1\rho_{\rm st}.
\end{equation}
From Figure~\ref{density-cont} we see that  the evolution of the  stars will be
dominated by  gravitational interactions with the BHs -- as oppose to self-interactions -- within 
a radius of size $\sim 3$pc, roughly the MBH influence radius. We conclude that the post-infall 
long-term evolution of 
the systems presented here will be very different from that of the models discussed
 in Section~\ref{sec-mass-seg} for which about one Hubble time was required in order to first 
met the condition Equation~(\ref{bh-dom}).
  
 \begin{figure*}
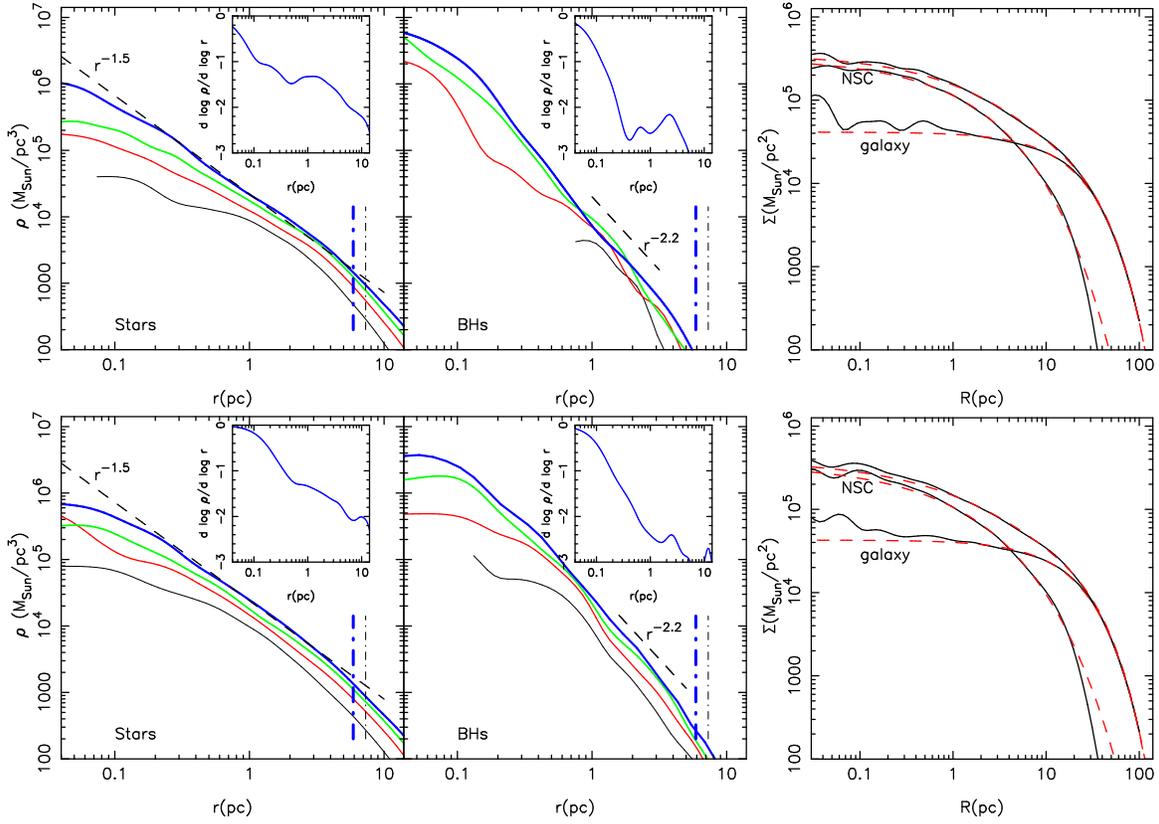
 
\centering
 \includegraphics[width=0.3\textwidth,angle=270]{Figure11a.ps} \includegraphics[width=0.3\textwidth,angle=270]{Figure11b.ps} \\
 \includegraphics[width=0.3\textwidth,angle=270]{Figure11c.ps} \includegraphics[width=0.3\textwidth,angle=270]{Figure11d.ps}
\caption{Density profile of the NC in  simulations B (upper panel) and C (lower panel) during the 
globular cluster inspirals. Left panels give the density of the NC (i.e., only particles that were initially associated with the clusters are used)  after 3, 6, 9 and 12 infall events (from bottom to top), while
the middle panels show the corresponding density distribution of BHs. Dot-dashed curves give  the radius containing 
a mass in stars twice the mass of the central MBH for the initial model (black-dot-dashed curve), 
and for the simulation end-products (blue-dot-dashed curve). Insert panels show the radial dependence of
the density profile slope $d \log \rho / d \log r$ in the final NC models.
Right panels display the projected density profile of  the stars in the NC,
in the galaxy (i.e., only particles that were initially associated with the galaxy) and the sum of them (upper curves).
Dashed-red curves are the best fitting  S\'ersic profile models to these projected distributions.}\label{DEN1}
\end{figure*}

 \begin{figure} 
\centering
 \includegraphics[width=0.3\textwidth,angle=270]{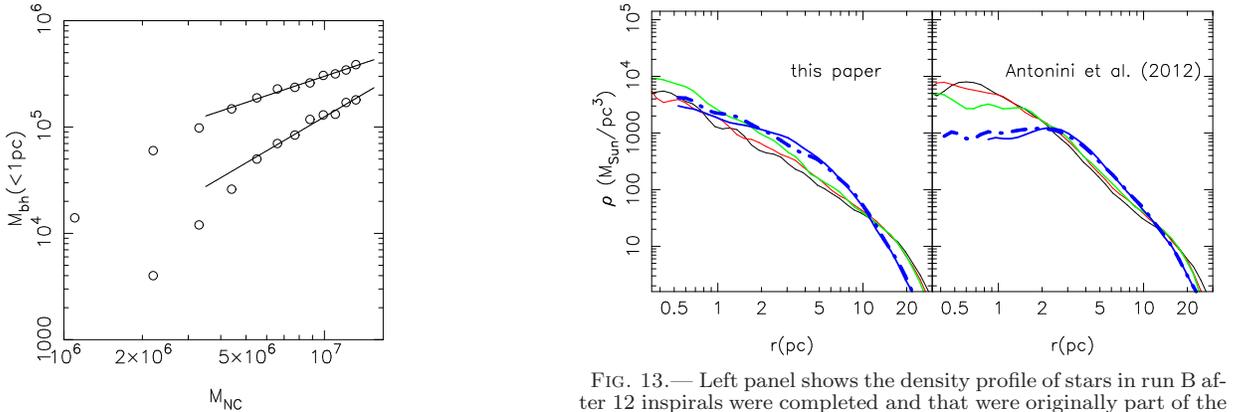} 
 \caption{Total mass  of BHs accumulated inside the inner parsec during the inspiral simulations  B (lower points) and 
 C (upper points), as a function of the  mass of the NC (i.e., the sum of the decayed globular cluster masses). 
 The solid curves give the best fitting power law profiles to the data for $M_{\rm NC}>4\times 10^6M_{\odot}$. Masses are 
 in units of solar masses.
 }\label{mbh_mnc}
\end{figure}

 \subsection{Consecutive inspirals} \label{multiple}
In order to determine the distribution of stars and stellar remnants predicted by a dissipationless 
formation model for the Milky Way NC we performed simulations which followed the repeated merger
of mass-segregated massive clusters in the GC; these correspond to simulations B and C of Table 1. In these integrations 
the number of $N$-body particles was much reduced with respect to the single inspiral simulations
presented in Section~\ref{single}. The subsampling was a necessary compromise to keep 
the computational time from becoming excessively long, while still allowing the simulations to follow
the successive inspiral of 12 clusters into the center of the galaxy, giving a total accumulated mass in stars 
 of $\approx 10^7~M_{\odot}$, which is roughly the observed mass of the Milky Way NC.

 The process of NC formation is illustrated in Figure~\ref{DEN1} which shows the growing central density  of stars and BHs  during the globular cluster inspirals. As clusters merge to the center, the peak density 
 of the model increases and  a NC forms, appearing within the model inner $\approx 10~$pc as an excess density of stars over
 the   background density of the galaxy. Insert panels give the radial dependence of the density profile slope of
 the final NC models.
 In the radial range $0.5~{\rm pc} \lesssim r \lesssim 5~$pc the 
 spatial density profile of the merger product  is characterized by  a density of stars which rises 
 steeply toward the center roughly as $\sim r^{-1.5}$. At smaller radii at the end of both simulations B and C the 
 spatial density  profile of stars flattens  and the radial dependence becomes approximately 
 ${ \rm d} \log \rho/{\rm d} \log r\approx -1 $ inside $0.3-0.5~$pc.  
 The BH population exhibits a very steep 
 density cusp ${ \rm d} \log \rho/{\rm d} \log r\approx -2.2$,
outside $\sim 0.5~$pc, and a somewhat flat profile within this radius.
The merger process produces a NC which is in an state of advanced mass-segregation, 
 with the heavy component dominating the density of stars inside a 
 radius of roughly $0.3~$pc and $1~$pc in simulations B and C respectively. 
 Since smaller systems   have shorter relaxation times
and undergo mass-segregation more quickly,  the merger process effectively reduces 
 the mass-segregation time scale of the NC compared for instance to the models
 discussed in Section~\ref{sec-mass-seg}.

\begin{figure}
\centering
 \includegraphics[width=0.26\textwidth,angle=270]{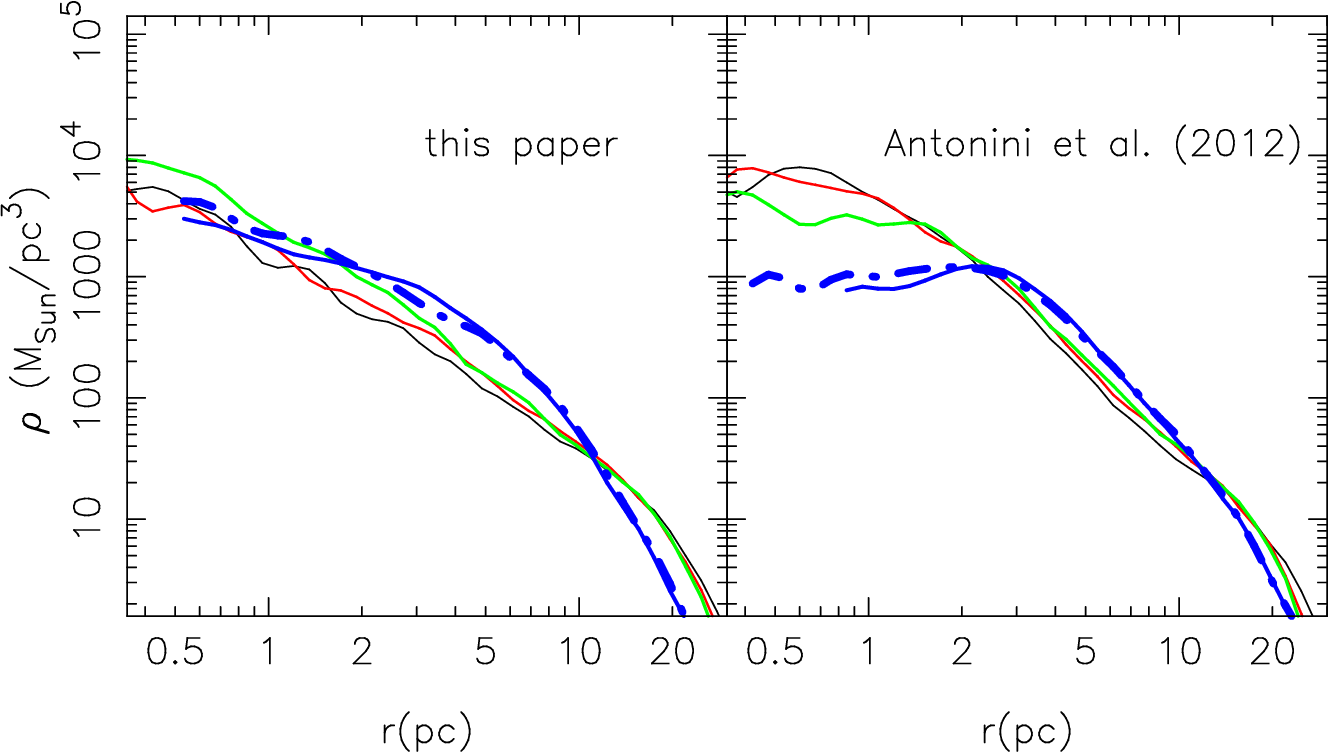} \\
\caption{Left panel shows the density profile of stars in
 run B after 12 inspirals were completed
and that were originally part of  the 3th, \textcolor{red}{6th}, \textcolor{green}{9th} and \textcolor{blue}{12th} infalling cluster. Right
panel corresponds to the single mass component simulations of \citet{AM12-2}.
Blue-dot-dashed curves show the density profile of stars coming from the 12th decayed cluster after
the final NC model was run in isolation for a time corresponding to roughly the time that takes in our simulations
for three consecutive inspiral events to complete.}\label{DEN2}
\end{figure}

The right panels of Figure~\ref{DEN1} show the projected density of the $N-$body model at 
the end of the simulations. 
These profiles were fitted as a superposition of two model components, 
one  intended to represent the galaxy and the other the NC.
For both components we adopted the S\'ersic law profile:
\begin{equation}\label{eq:sersic}
\Sigma(R)=\Sigma_{0} \text{exp}{\left[ -b\left( \frac{R}{R_{0}} \right)^\frac{1}{n} +b\right]}~,
\end{equation}
with
\begin{equation}
b=2n - \frac{1}{3} + \frac{0.009876}{n}.
\end{equation}
For the end product $N$-body model of run B, the best-fit parameters were 
$\Sigma_{0}=7.21\times10^3~\text{M}_\odot/\text{pc}^2$,
$n=1.03$, $R_{0}=31.9$~pc  for the bulge, and 
$\Sigma_{0}=1.03\times10^4~\text{M}_\odot/\text{pc}^2$, $n=1.88$, $R_{0}=9.51$~pc
for the NC. The best-fit parameters of the merger product of run C were
$\Sigma_{0}=7.44\times10^3~\text{M}_\odot/\text{pc}^2$,
$n=1.04$, $R_{0}=31.6$~pc  for the bulge, and 
$\Sigma_{0}=7.60\times10^3~\text{M}_\odot/\text{pc}^2$, $n=2.08$, $R_{0}=11.3$~pc for the NC.
 
We note that the final density profile and structure of the NC depends on a variety of factors, 
 these include  the initial distribution of stars and BHs  inside
the clusters,  the strength of the tidal field due to the galaxy and MBH, 
and how the distribution of previously migrated
stars and BHs evolves in response  to their gravitational interaction with other
 background stars and with infalling clusters.  For instance, the initial degree of internal evolution 
 in the cluster models together  with the adopted galactic MBH mass will determine how close 
 to the galactic center the BH  and  stellar clusters will get before they completely dissociate; in turns
 this regulates the size of the region over which the stellar distribution 
 flattens (i.e., the core size of the NC density profile), as well as 
 the number of stellar BHs transported in the vicinity of the MBH.

In Figure \ref{mbh_mnc} the mass in BHs accumulated in the inner parsec, $M_{\rm bh}(<1{\rm pc})$, is
shown as a function of the NC mass, $M_{\rm NC}$. At any time the NC mass is given 
by the sum of the accumulated globular cluster masses.
A good fit to the data for $M_{\rm NC}>4\times 10^6M_{\odot}$ is given by: 
$M_{\rm bh}(<1~{\rm pc})=a\times (M_{\rm NC}/{4\times 10^6M_{\odot}})^b$, with 
$a=3.33\times 10^4 M_{\odot}$ and $b=1.44$ for Model B, and 
$a=1.42\times 10^5 M_{\odot}$ and $b=0.819$ for Model C; these fitting functions are 
plotted in Figure \ref{mbh_mnc} as solid curves.

The effect of the infalling clusters on the pre-existing NC is  illustrated in the 
left panel of Figure~\ref{DEN2} which shows the density profile of stars in
 simulation B after 12 inspirals were completed
and that were originally inside the 3th, 6th, 9th and 12th infalling cluster. 
The density profile of  stars from the 12th merged cluster,
consistently with the results of the high resolution 
simulations  of Figure~\ref{density-single-inpiral}, is characterized by a 
shallow density cusp out to roughly $1-2~$pc  and an outer envelope with density
that falls off rapidly with radius. The dot-dashed curve in the figure illustrates the density profile 
of the same stars after the final NC model was run in isolation for a time equal
to the time for three consecutive inspiral events to occur, $\sim 5\times 10^7$yr. The similarity between the two 
mass distributions indicates that collisional two-body relaxation, due to random star-star and star-BH
gravitational  encounters, can be ignored during the inspiral simulations.
The densities  of stars coming from previously decayed globular clusters, however, 
appear very similar to each other and
very different from the density  of stars transported during the last inspiral, 
being  lower  and having a steeper
radial dependence, ${ \rm d} \log \rho/{\rm d} \log r\approx -1.5$, in the radial range 
$1~{\rm pc} \lesssim r \lesssim 10~$pc.

\begin{figure}
 \includegraphics[width=0.35\textwidth,angle=270]{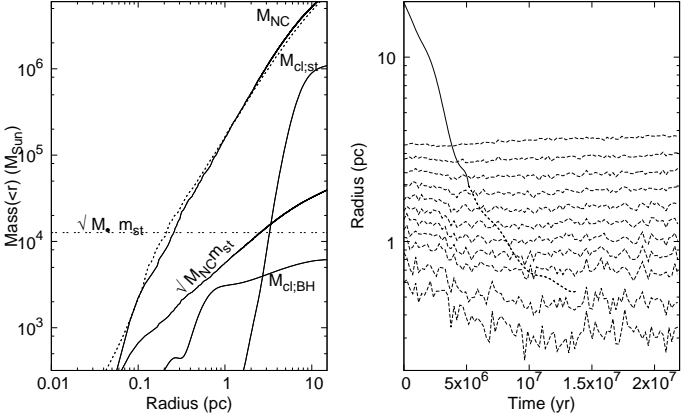}\\
 \caption{
Left panel:
  upper curves display the cumulative mass distribution of stars  ($M_{\rm NC}$) in the NC
  at the beginning (solid curve) and at the end (dotted curve) of the 12th inspiral.
   The total mass in stars ($m_{\rm cl,st}$)   and BHs ($m_{\rm cl,bh}$)  as a function of 
   galactocentric radius of the 12 inspiraling cluster are displayed; 
   time increases from right to left  as the cluster orbit decays 
   toward the center and loses mass.
 The value of $\sqrt{M_{\rm NC}m_{\rm st}}$ is also illustrated as a function of radius;
 this is roughly equal to $\sqrt{M_{\bullet}m_{\rm st}}$ (double-dotted  curve) at the MBH 
 radius of influence. 
 When the total cluster  mass is larger than $\sqrt{M_{\rm NC}m_{\rm st}}$, collisional relaxation in the 
 NC is dominated by scattering of stars and BHs from the infalling cluster.
 In the right panel we display the Lagrangian radii of stars from the previously decayed (11th) cluster
 during the 12th inspiral (dashed curves), 
 and the evolution of the orbital radius of the 12th cluster. The orbit of the cluster is shown using a continue
 curve when $m_{\rm cl} > \sqrt{M_{\rm NC}m_{\rm st}}$, and a dotted curve otherwise.
 The evolution of the Lagrangian radii 
 clearly shows that the inspiral modifies the mass distribution of the pre-existing NC, as argued in the text.
   }\label{mvsr2}
\end{figure}

As illustrated in the right panel of Figure~\ref{DEN2},
this evolution was also present in the one-component inspiral simulations of \citet{AM12-2};
the initial conditions of these simulations were essentially the same as those of simulation B with the
difference that the clusters only had a single mass particle group representing stars.
There are two mechanisms that drive the evolution of the density profile toward their final form: (1)
stars from earlier infalling clusters are stripped at smaller radii and dominate inner regions of
 the NC~\citep{hagai+ale} 
and, as argued in the following section,  (2)  gravitational scattering
of the previously accumulated stars and BHs by the infalling clusters.

\begin{figure}
\centering
$\begin{array}{ll}
  \includegraphics[width=0.35\textwidth,angle=0]{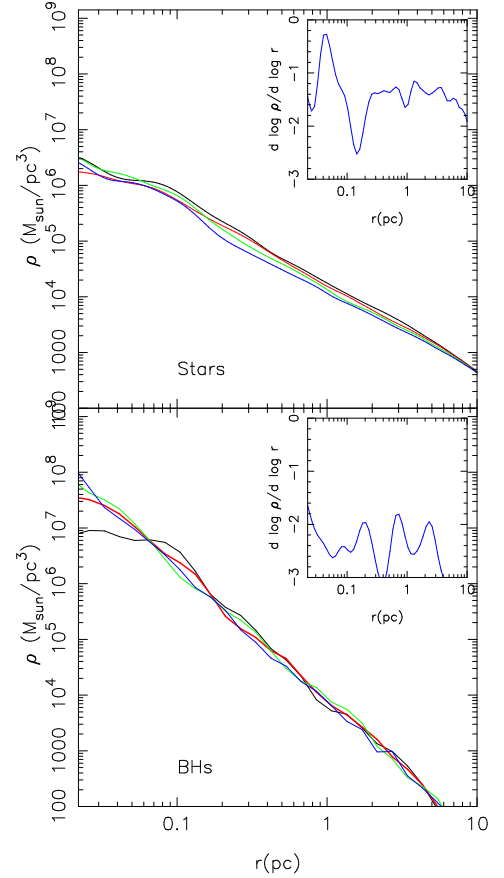}
\end{array}$
\caption{Evolution of the density profile of stars and BHs during the post-merger simulation of model B at
  $t=(0.09,~\textcolor{red}{0.18},~\textcolor{green}{0.27},~\textcolor{blue}{0.36})\times T_{r_{\rm infl}}$.
  Insert panels give the radial dependence of the density profile slope $d \log \rho/d \log r$ of
  the final model ($t=0.36~{\rm T_{r_{\rm infl}}}$).
  }
  \label{msclu}
\end{figure}

 \subsection{Accelerated cusp regrowth due to scattering from infalling clusters}\label{acc-rel}
Consider   a massive cluster of mass $m_{\rm cl}$ which moves into a system of $N$ stars.
The second order diffusion coefficients 
that appear in the Fokker-Plank equation, and which describe the evolution
of the stellar distribution due to self-scattering ($\mathcal{D}_{EE,11}$) and 
scattering off of the cluster ($\mathcal{D}_{EE,1~ {\rm cl}}$),  
scale with the mass of the perturber, the mass~($m$) and density of field particles as:
\begin{eqnarray} \label{dc}
\mathcal{D}_{EE,11} \simeq  m^2 N; ~~
\mathcal{D}_{EE,1~ {\rm cl}} \simeq m_{\rm cl} ^2 ,
\end{eqnarray}
where here we have approximated
the cluster as a point mass perturber.
From Equation~(\ref{dc}) we see that if $m_{\rm cl}\gg \sqrt{m^2N}$,
$\mathcal{D}_{EE,1 ~{\rm cl}}\gg \mathcal{D}_{EE,11}$, i.e.,
self-scattering is negligible compared with scattering off of the massive
perturber; the first order coefficients can also be ignored since they
are smaller than $\mathcal{D}_{EE,1~{\rm cl}}$ by factors of $m^2N/m_{\rm cl}^2$
and $m/m_{\rm cl}$. 
Thus, if 
\begin{equation}\label{cond-cl}
m_{\rm cl}> \sqrt{m^2N}
\end{equation}
an inspiraling cluster will reduce the energy
relaxation time scale compared to that due to stars alone
by a factor $\sim m_{\rm cl}^2 \ln \Lambda /m^2N \ln \Lambda' $. 
In the latter expression, $\Lambda '$ represents the Coulomb logarithm estimated by taking into account the
large physical size of the cluster, which implies a lower effectiveness of close gravitational scattering
with respect to star-star scattering~\citep[e.g.,][]{Merritt-book}. This latter quantity can be set to 
\begin{equation}
\ln \Lambda ' \approx \frac{1}{2} \ln \left[ 1+ \frac{p_{max}^2}{p_0^2}\right]
\end{equation}
with $p_{max}$ roughly 1/4 times the linear extent of the test star's orbit. Setting this size to $5~$pc, and
$p_0$ to half of the size of the cluster, $\approx 1~\rm pc$, the typical relaxation time becomes:
\begin{eqnarray}\label{teff}
T^{eff}_r(r)\approx \frac{0.34~\sigma^3(r)}{G^2  \ln{\Lambda'} \rho} \frac{M_{\bullet}}{m_{\rm cl}^2} 
= 2 \times 10^7  \left( \frac{\sigma}{50 {\rm km s^{-1}}} \right)^3 \\
\times \left(\frac{1.6}{\ln \Lambda'} \frac{10^3 M_{\odot}}{\rho} \frac{M_{\rm \bullet}}{4\times 10^6~M_{\odot}}\right)
\left(\frac{5\times 10^5~M_{\odot}}{m_{\rm cl}}\right)^2   {\rm yr}  \nonumber
\end{eqnarray}
where in deriving this expression we have used the fact that  
$m N \approx M_{\bullet}$ near the sphere of influence of the MBH.
 For $M_{\bullet}\sim 10^7~M_\odot$, the infall of even a cluster of $10^4~M_\odot$ would largely 
affect the rate of collisional relaxation, provided that the cluster spends a time of order $T^{eff}_r$ in a region
where the condition~(\ref{cond-cl}) is satisfied.

 The left panel of Figure~\ref{mvsr2} shows the total mass of  BHs and stars
bound to the 12th infalling cluster in run B as a function of cluster orbital radius
and compares these to  $\sqrt{m^2N}$. The cluster spends  roughly
$5\times 10^6~$yr before it reaches $\sim 2~$pc (right  panel of Figure~\ref{mvsr2}) 
after which its mass drops below $\sqrt{M_{\bullet} m}$
and self-scattering starts to be the dominant effect driving collisional relaxation.  Since
the time for the cluster
to reach this radius is comparable to $T^{eff}_r$, the inspiral of clusters  
is expected to have an important impact on the density distribution of the preexisting NC, in agreement with the results of 
our simulations. In the right  panel of Figure~\ref{mvsr2} we show the evolution of the Lagrangian radii 
of the stars transported in the NC during the previous (11th) inspiral. During the first   
$10^7~$yr, the condition (\ref{cond-cl}) is satisfied and the inspiral induces a rapid evolution
of the NC mass distribution.
Note that in a real galaxy, due to the smaller individual stellar mass,
scattering from the perturber would dominate down to smaller radii than in our $N$-body model.
However,  in the region where  the condition~(\ref{cond-cl}) is satisfied the relaxation time
 in the simulations would be roughly the same as in the real system.
 
 Scattering of stars by the massive perturber will cause an initial  density core
to fill up and the distribution function to  evolve toward a constant value. The result is a sharp increase
in the density profile of stars, $n\sim r^{-1.5}$,
inside a radius approximately  equal to the radius within which the condition Equation~(\ref{cond-cl})
is first met. After the  perturber reaches a radius, $r_{\rm crit}$,  containing a total mass in stars
smaller than its mass, a large density core is rapidly carved out 
as a consequence of ejection of stars from the center. This is the 
evolution that for example characterizes  the mass distribution of stars during the formation and evolution of MBH binaries 
in galactic nuclei~\citep[see Figure~13 and 14 of][]{AM12}. In our simulations, however,
the clusters disrupt before  reaching $r_{\rm crit}$, and before the second phase of cusp disruption can
initiate. Thus, the net effect of infalling clusters on the pre-existing NC distribution is that of inducing a short period of enhanced 
collisional relaxation, which causes the central density of stars and BHs to increase during the inspirals.

 \subsection{collisional evolution}\label{relax}
 During and after its formation a NC will evolve due to collisional star-star, star-BH, and BH-BH interactions which will cause 
its density distribution to slowly morph into the steeply rising density profile
which describes the quasi-steady state solution of stars and BHs near a MBH.
We study the evolution of the NC due  to collisional relaxation by evolving the inspiral simulations end-product 
of simulation B for roughly half a relaxation time (or one Hubble time when scaling the $N$-body 
model to the Milky Way). 
In order to efficiently evolve the system for such a long timescale  the 
$N$-body model was resampled to contain a smaller number of particles. Since we are interested in the 
model distribution at small radii  we kept the mass of the particles the same as in the original model, so that the  resolution 
of the simulation in the region near the center was unchanged, and we
included only particles with orbital periapsis 
less than $10$pc and apoapsis less than $20$pc.
In this way the total gravitational force acting on particles lying inside $\approx 10$pc was approximately unchanged
with respect to the original model.
This sampling  procedure is  equivalent of  truncating 
the  mass distribution  of the model smoothly at $r\gtrsim10$pc. $N$-body integrations 
over a  few crossing times verified the (collisionless) quasi-equilibrium state of the truncated model.
 
 Since the dynamical friction times for globular clusters at $20~$pc from the galactic center 
are much shorter than relaxation times,  both  cluster and 
 galaxy will not undergo a significant amount of collisional relaxation during the inspirals. 
 In our $N$-body simulations B and C the inspiral time of the globulars to reach the center 
 and disperse around the central MBH is about $100$ times shorter than the relaxation time of the NC.  
 Thus, during the inspiral simulations, relaxation due 
 to star-star, star-BH, and BH-BH interactions was ignored,
while in the  post-merger simulations presented here times were scaled to 
the relaxation time computed at the sphere of influence of the MBH.
Note that for the sake of simplicity the evolution was broken
into two successive stages: infall of the clusters; then evolution, due to two-body encounters, 
of the stellar distribution around the MBH with infalls ``turned off.'' In reality, subsequent inspirals 
would be separated by times of order a Gyr and significant two-body relaxation would occur between these events.
  
 Figure~\ref{msclu} shows the evolution of the  mass density in each species in the post-merger integrations. 
Initially, the density of BHs dominates over the density in stars at $r\lesssim 0.1r_{\rm infl}$. Thus, at radii smaller than these,
the evolution of the BHs is mostly  driven by BH-BH self-scattering which causes
their distribution to reach,  in roughly $0.2\times T_{r_{\rm infl}}$,
a quasi-steady state form  characterized by a steep density slope, $\gamma \approx 2.2$,
at all radii. At radii larger  than $\sim0.1r_{\rm infl}$ the stars dominate the mass 
density and the BH population evolves  due to dynamical 
friction against the lighter component. The general trend is for the central mass density of BHs to
steadily increase with time.

While the  BHs segregate to the center, the lighter particle mass density 
 decreases  within the radial range $0.1\lesssim r\lesssim 10~$pc.
The evolution of the stellar distribution toward lower densities is a consequence of their gravitational 
interaction with the BHs:
when the density in the BHs approaches locally the density in stars, 
heating of the light particles off the heavy particles dominates over star-star scattering, 
causing the density of the former to decrease. 
Scattering of stars off the BHs is also expected to promptly modify the 
star density profile  at small radii causing the formation of a ``mini-cusp" at 
$r\lesssim 0.02r_{\rm infl}$~\citep{GM12}.  We find  evidence of this phenomenon 
in our $N$-body models which rapidily develop a $n\sim r^{-1.5}$ 
cusp at $r<0.03~$pc~(see upper panel and corresponding insert panel of Figure~\ref{msclu}).
 Outside these radii however, even after about half a relaxation time the star distribution retains
the shallow density profile at $r\lesssim 0.2~$pc that characterized the initial model.
We conclude that even after a time of order the  relaxation time, model B looks 
different from the dynamically relaxed  models that
are often assumed to describe the density distribution of
stars and BHs near the center of galaxies.
After $\approx 0.4\times T_{r_{\rm infl}}$  the BHs (stars) attain a 
central density cusp which is steeper (shallower) than in those models.

 We note that the details  of the final density distribution of our  NC model
depends on a number of factors that remain quite uncertain. For example, the final state of the BH population in the NC
and the degree at which the BHs are segregated in the final model will depend on  their initial number fraction
and in turns on the assumptions made for  their initial distribution in the parent clusters.

\begin{figure}
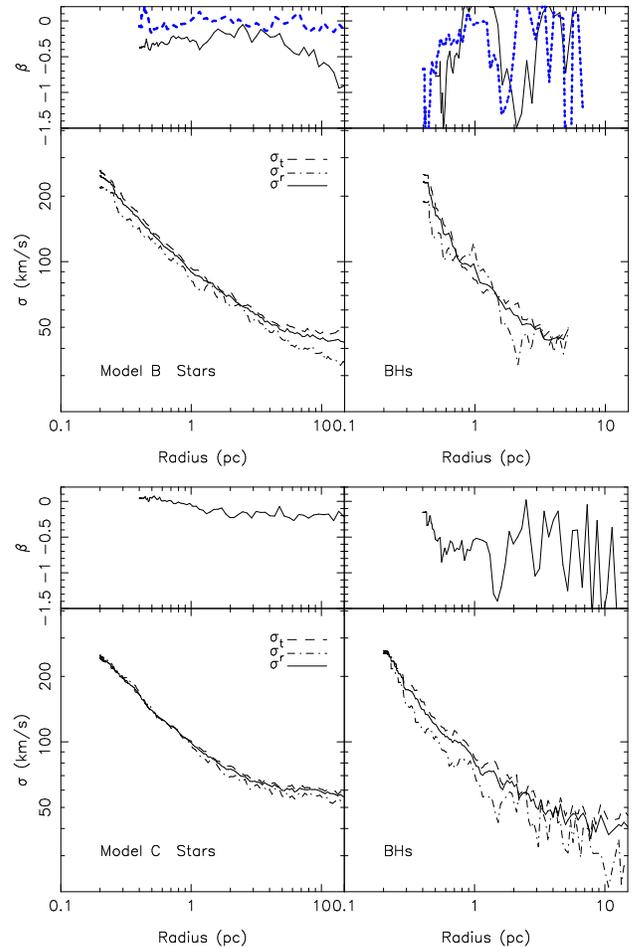

\centering
$\begin{array}{ll}
  \includegraphics[width=0.342\textwidth,angle=270]{Figure16a.ps} \\
   \includegraphics[width=0.34\textwidth,angle=270]{Figure16b.ps} 
\end{array}$
\caption{ Radial profile of the anisotropy parameter ($\beta$) and
 radial ($\sigma_r$) and tangential ($\sigma_t$) components of 
the one-dimensional velocity dispersion~($\sigma$). Plots were obtained  
at  the end of the 12th inspiral event using only
$N-$body particles that were originally inside the clusters. 
The blue-dashed curves in the upper panels display the anisotropy profile 
of the NC in simulation B at the end of the post merger phase. 
}
  \label{vdisp}
\end{figure}

 \subsection{Kinematics and morphology} \label{k-m}
Understanding the kinematical structure and shape of galactic nuclei is important for placing
constraints on their formation history and evolution. 
We quantified  the velocity anisotropy  of our models by  using the parameter:
\begin{equation}
\beta=1-\frac{\sigma_t^2}{2\sigma_r^2}~,
\end{equation}
with $\sigma_r$ and $\sigma_t$ the radial and tangential velocity 
dispersions respectively. Figure~\ref{vdisp} shows  the radial profile 
of $\beta$, $\sigma_r$ and $\sigma_t$   for the
stellar and BH populations in simulations B and C.  { At the end of the inspiral simulations, model B 
is characterized by an approximately flat velocity anisotropy profile with $\beta \approx -0.5$ while in model C, $\beta$
slightly decreases from nearly $0$ to $-0.2$ within $1~$pc and it is approximately constant outside this radius.
Thus, our models are \emph{tangentially} anisotropic throughout $100~$pc, both in the stellar and BH components.
The upper panels of  Figure~\ref{vdisp} also display  the radial profile of the anisotropy parameter of model B at the
end of the post merger phase, i.e., after the system was run in isolation for a time of order the  relaxation time. 
Evidently, two-body relaxation causes $\beta$ to increase
and the NC to evolve toward spherical symmetry in velocity space.

\begin{figure}
\centering
$\begin{array}{ll}
\\  \includegraphics[width=0.4\textwidth,angle=270]{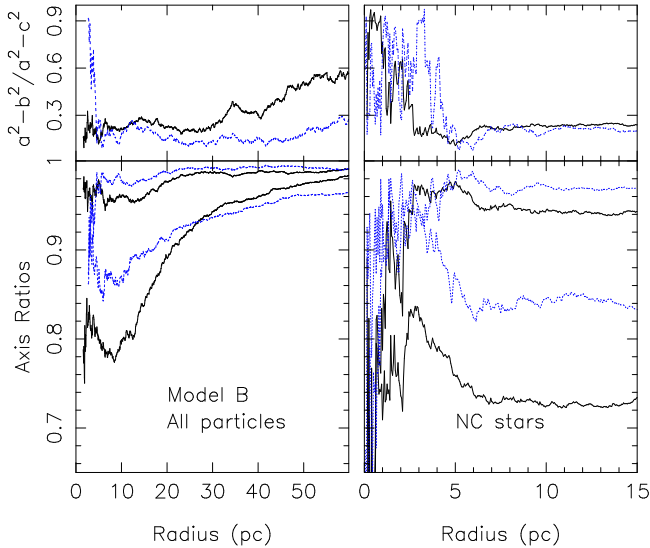} 
  \end{array}$
\caption{
 Triaxiality parameter (upper panel) and axis ratios (lower panel) as functions of radius, 
for model B at the end of the inspiral phase~(black-solid curves) 
and at the end of the post-merger collisional evolution~(blue-dotted curves).
Left panels give the shape parameters obtained by including all particles (stars and BHs)
in the evaluation of the symmetry axes; in the right panels only particles representing stars 
that were transported to the center by the infalling  stellar clusters were used.
}
  \label{shape}
\end{figure}

We measured the model shape in our simulations 
 from the moment-of-inertia tensor~\citep[e.g.,][]{Katz:1991,PM:04, ACM09}.
The  symmetry axes are calculated as
\begin{equation} \label{rapp_ax}
\tau_1=\sqrt{I_{11}/I_{max}}~,~\tau_2=\sqrt{I_{22}/I_{max}}~,~\tau_3=\sqrt{I_{33}/I_{max}}~,
\end{equation}
where $I_{ii}$ are the principal moments of the inertia tensor
and $I_{\mathrm {max}}={\rm max}\{I_{11},I_{22},I_{33}\}$;
particles are then enclosed within the ellipsoid $x^2/{\tau_1}^2+y^2/{\tau_2}^2+z^2/{\tau_3}^2=r^2$.
These previous steps were iterated until the values of the axial ratios  had a percentage change of less than $10^{-2}$.
We also define a triaxiality parameter: $T\equiv \left(a^2-b^2\right)/\left(a^2-c^2\right)$.
The value $T=0.5$ corresponds to the \lq maximally triaxiality\rq ~case, while
oblate and prolate shapes correspond to $T=0$ and $1$, respectively.

Figure~\ref{shape}  displays  radial profile of the axis ratios  and triaxility parameter of the model
in simulation B at the end of the inspiral simulation (black-solid curves) and at the end of the post-infall evolution (blue-dotted curves).  
We also evaluated the shape of the NC component by only including the stars that were transported to the center from the 
infalling stellar clusters (right panels); 
the NC component appears strongly triaxial-like within 5pc and mildly triaxial (quasi-oblate)  outside this radius.
The formed NCs in simulations C and B shared a similar morphological structure, 
so we only displayed results for model B.
The model morphology within $\sim 30$pc is initially mildly  triaxial  and evolves due to gravitational encounters
toward a more quasi-spherical shape.  
It is important that at the end of the post-merger phase 
the model still exhibits  a significant degree of triaxiality, $0.1\lesssim T \lesssim0.3$. 
In fact, such level of asymmetry might be large enough to significantly increase the number of tidal
captures of stars and stellar binaries  when compared  to the same rate obtained in 
collisionally resupplied loss cone theories where spherical geometry is often assumed both in configuration 
and velocity space~\citep{MV2011}.

\section{Discussion}\label{discussion}

\subsection{Comparing to  the properties of the Galactic Nuclear Cluster}

Star counts using adaptive optics spectroscopy and medium-band imaging have shown that the red giants 
at the Galctic center, the only old stars that can be resolved in these regions,
have a flat projected surface density profile close to Sgr~A*~\citep{BSE,D:09}.
The core in the red giant  population extends out to approximately $0.3-0.5~$pc from the center.
However, due to the effect of projection, it is difficult to constrain  core size and  three-dimensional spatial density profile,
which could be slowly rising but even declining toward the center.
 
It is possible that a  cusp in the lower mass stars is present and that the observed core
is the result of a luminosity function that changes within $0.5~$pc.
This could be due to physical collisions before or during the giant phase~\citep{bailey+davies99,dale+09}; 
or tidal interactions between  stars and the  central MBH~\citep{davies+king05} --
mass removal can make the luminosity that a star would otherwise reach at the tip of the red-giant phase considerably fainter and
prevent  these objects from evolving to become observable. 
While these models are possible  they seem not to fully explain the observations~\citep{dale+09}. Thus, it is 
important to consider the possibility
 that the  red giants are indeed representative of the unresolved low-mass main-sequence stars, and that  
the the density core is a consequence  of  the NC formation history combined to a 
long nuclear relaxation timescale~(Section~\ref{sec-mass-seg}).
On this basis we can directly compare the predictions of theoretical models to the kinematics 
and mass distribution inferred from observations of the giant stars at the GC and  
draw conclusions about possible formation  mechanisms for the Milky Way NC.

\subsubsection{Mass distribution}
Figure~\ref{DEN1} shows that the merger of about $10~$clusters results into a compact nucleus with a stellar density profile that declines
as $n \sim r^{-1.5}$ outside $\sim0.5~$pc and  flattens to $n \sim r^{-1}$ within this radius. 
More precisely, we consider two definitions of the core radius: (1) the projected radius, $r_c$, 
at which the surface density falls to one-half its central value; (2) the break radius, $r_b$, at which 
the density profile transits from the inner law~($n \sim r^{-1}$) to the outer density law~($n \sim r^{-1.5}$).
We find $r_c = 0.51$pc and $r_b=0.48$pc for Model~B, and $r_c = 0.61$pc and $r_b=0.59$pc for Model~C.
Collisional relaxation occurring during the post merger simulation of Model~B reduces the size of the core with time, while
inner and outer density slopes remain roughly unchanged.
At the end of the simulation, after $\sim 0.3T_{\rm r_{\rm infl}}$,  we find $r_c = 0.32$pc and $r_b= 0.18$pc. 
The size of the region where the density  transits to a shallow profile in our models appears to be 
comparable to the  extent of the core inferred from number counts of the red giant stars at the GC.
{ The exact extent of the core region (and how far our models are from their steady state) is determined by a number of factors that depends on the assumptions made in the $N-$body initial conditions. For example, the size of the density core could be made larger or smaller depending on the initial degree of cluster evolution and the number fraction of BHs.  
However,  we note that the presence of a core in the final density distribution seems to be a
quite robust outcome of a merger  model for NCs. For example,
the single mass component simulations of \citet{AM12-2} also produced a final NC model with a 
core of size $\sim 1$pc, somewhat  similar to what we find in this paper. 
The absence of a Bahcall-Wolf cusp is naturally 
explained in these models, without the need for fine-tuning or unrealistic initial conditions.

Our simulations result in a final density profile having nearly the same power-law index beyond~$0.5$pc
as observed~\citep[$\Sigma(R)~R^{-1}$,][]{haller+96}. The slope index in the inner $\sim 0.5~$pc of our model, $\gamma \sim 1$,
 is also consistent with what obtained from the surface brightness 
distribution of stellar light within the inner $1''$ of Sgr A*~\citep{yz+12},
but appears only  marginally consistent  with what inferred from number counts of the red giant stars.
Slope indexes in the range ~$ -3 \lesssim \gamma \lesssim 0.8$
are consistent with what derived from observations of the giants, 
although negative or nearly-zero values, corresponding to centrally-decreasing  or flat densities respectively,
are preferred \citep{M:10,do+13}.

 \subsubsection{Kinematics}
 The radial profile of the velocity anisotropy of the NC could potentially provide useful constraint on its formation. 
 Kinematic modeling of proper motion data derived from the dominant old population of giants, reveals a nearly spherical 
 central cluster exhibiting slow, approximately solid-body rotation, of amplitude $1.4~{\rm km s^{-1}arcsec^{-1}}$~\citep{trippe,S09}.
 Kinematically, the central cluster appears isotropic, with a
mildly radial  anisotropy at $r\lesssim 0.1~$pc and slightly tangentially
anisotropic for $0.1~{\rm pc}\lesssim r\lesssim~1~{\rm pc}$;  
In the radial range $1''-10''$, the late-type stars are observed to have a mean projected anisotropy of
$\langle 1-{\sigma_T^2/\sigma_R^2\rangle=-0.12^{+0.098}_{-1.05}}$~\citep{S09}.
\citet{do+13} found $\beta=0.01^{+0.35}_{-0.34}$ within $0.5~$pc.

Our models are characterized a generally flat anisotropy profile with $\beta \approx -0.5$ at the end of the inspiral simulations
and $\beta \approx 0$ at the end of the post-merger evolution. Although such values are consistent with observations, 
we believe that future and better kinematic data that extend outside the inner parsec will be necessary in order 
to provide better constraints on this scenario.
 
We note that due to our assumption of no preferential direction of inspiral, the merger remnants in our simulations showed no significant net rotation.
Recent observations of the NC in our Galaxy suggest instead a significant rotation on parsec scale \citep{S14}; 
this might be reconciled with a cluster merger origin for the Galactic NC if, for instance, the clusters 
were originally dragged down into the Galactic disk plane (where they experience an a greater dynamical friction force) and transported into the
central region of the Galaxy where they then accumulated to form
a dense nucleus which will then appear to rotate in the same sense of the Galaxy.

\subsubsection{Kinematically cold sub-structures}
\citet{Feldme+14}  found indications for a substructure in the Galactic NC that is rotating approximately perpendicular to the Galactic rotation with $\sim30~{\rm km~s^{-1}}$ at a distance of $\sim20"$ or $0.8~$pc from Sgr A*. In addition they found an offset of the rotation axis from the photometric minor axis and argue that this hints to infalling clusters. 

We look for kinematic substructures in our models by using  the Rayleigh \citep[dipole,][]{rl-19} statistics $R'$ defined as the length of the resultant of the unit vectors $l_i,~i=1,...,N,$ where $l_i$ is perpendicular to the orbital plane of the $i^{th}$ particle and $N$ is the number of particles  (a total of 5715 per cluster). For each merged cluster we computed $R'$ over the entire course of the inspiral simulation. For a fully isotropic distribution we expect $R' \sim \sqrt{N}$, while $R' \sim N$ if the orbits are correlated. 

As an example, Figure~\ref{ksub} gives  $R(=R'/N)$ as a function of time (scaled to the relaxation time of a Milky Way like nucleus).  Initially, after a cluster reaches the center the orbits are strongly correlated and, as expected, $R\sim1$, i.e., the stars from the infalling clusters  distribute into a thin disk configuration initially. Due to two-body relaxation, and due to the perturbing effect of the later infalling clusters, $R$ decreases  with time and approaches a value more consistent with isotropy. 
Figure~\ref{ksub} shows that all clusters maintain some degree of anisotropy during the entire corse of the simulation. But the orbits of  the stars from the first 7 clusters are almost completely isotropic by the end of the simulation.
The orbits of stars from the last four/five infalling clusters are still largely correlated after $\sim 3~$Gyr of evolution. 
A linear fit to the data 
($R$~vs~$t$) for  the last decayed cluster gives: $R(t) = -0.11\times t/{\rm Gyr}+0.96$, so that it would take about $\sim 10$~Gyr for the stars to achieve a nearly isotropic distribution. These results are consistent with those of \citet{mp13}, who found that it takes a time of order the relaxation time of the nucleus to fully randomize an initially cold disk.

\begin{figure}\centering
 \includegraphics[width=0.355\textwidth,angle=270]{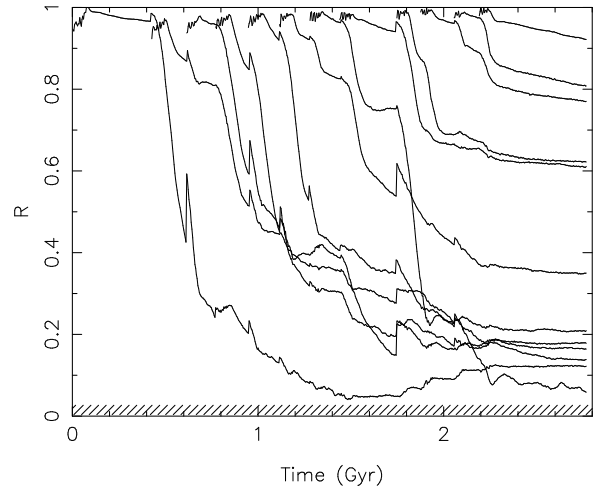}   
 \caption{ Evolution of the Rayleigh parameter $R=R'/N$ that measures the
 degree of randomness of the orbital orientation of  stars transported to the center by each of the 12
infalling clusters of model B.  The hatched region shows the $90\%$ confidence bands expected for a fully random distribution of orbital orientations of 5715 stars. At the end of the 12th inspiral event significantly cold kinematic
  substructures are present in the NC model.
  }
 \label{ksub}
\end{figure}

\begin{figure}\centering
 \includegraphics[width=0.355\textwidth,angle=270]{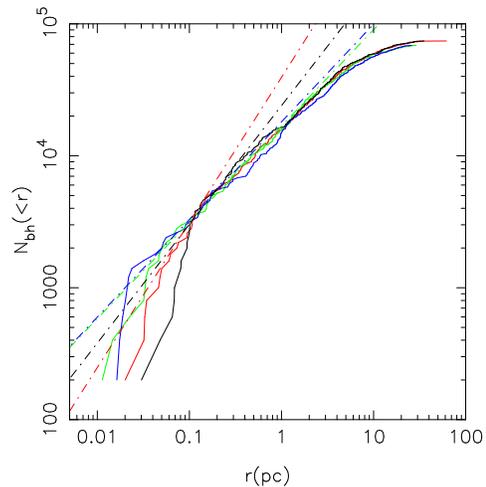}   
 \caption{Cumulative number of BHs predicted for a Milky Way nucleus as function of radius
  and at the different time 
  intervals: $t=(0.09,~\textcolor{red}{0.18},~\textcolor{green}{0.27},~\textcolor{blue}{0.36})\times T_{r_{\rm infl}}$,
  for the post-merger simulation discussed in Section~\ref{relax}. 
  Dot-dashed curves were obtained 
  by carrying out regression fits of $\log N_{\rm bh}$ to $\log r$ in the radial range 
   $(r_1 , r_2 )$, with $N_{\rm bh}(< r_1) = 5$  and $N_{\rm bh}(< r_2) = 25$ (see text for details).
  }
 \label{msclu2}
\end{figure}

\subsection{Extreme-mass-ratio inspirals}
The inspiral of compact remnants into a MBH represents one of the most promising
sources of gravitational wave radiation detectable  by space-based laser interferometers~\citep{AS12}. 
Event rates for such extreme mass-ratio  inspirals are generally estimated under the assumption 
that the BHs had enough time to segregate and form a steep central cusp, $n \sim r^{-2}$, near the MBH. 
Such dynamical models  predict inspiral rates per galaxies of $\sim 250~{\rm Gyr^{-1}}$~\citep{HA06}.
Models that include an initial parsec scale core can result in much lower central BH
densities than in the steady state models, and imply rates as low as $1-10~{\rm Gyr^{-1}}$~\citep{M:10}. 
EMRI event rates could be also severely suppressed by the 
Schwarzschild barrier which limits the ability of stars to diffuse 
to high eccentricities onto inspiral orbits~\citep{MAMW2011}.  

It must be stressed that it is difficult to draw any conclusion about EMRI event rates from the models
discussed in the literature  because of the significant uncertainties in the underlining assumptions. 
Our simulations cast  further doubts on results obtained from idealized time-dependent models which relay on the assumption that  BHs
and stars  have initially the same spatial distribution. For example, based on the models discussed in Section~\ref{sec-mass-seg} and 
in \citet{M:10} and \citet{AM12}, the presence of a core in the old population of stars at the GC would 
imply very low central densities of BHs and EMRI event rates. 
However, the initial conditions adopted in the simulation of Section~\ref{sec-mass-seg} 
 were quite artificial and not motivated by any specific physical model.
 We have shown that in a merger model for the formation of NCs, the resulting distribution of stellar remnants
partially reflects their distribution in their parent clusters just before they reach the center of the galaxy.
 Thus, different, possibly more realistic, initial 
conditions would produce rather different central BH densities;  in these models EMRI rates could be as large as (or higher than)  those  
obtained in the steady state models even in the presence of a core in the stellar distribution.

In order to determine the number of BHs in a Milky Way like nucleus predicted by the inspiral simulations
we used the scaling relation given by Equation~(\ref{bh-sc}). 
We are interested in the number of compact remnants inside $\approx 0.01$~pc, as these
are the only BHs that can generate EMRIs. 
Since  at these radii the number of BH particles  in the $N-$body model  is small, 
we carried out regression fits of $\log N_{\rm bh}$ to $\log r$ at larger radii,
and extrapolate inward in order to get $N_{\rm bh}$ at the radii of interest. 
Following \citet{GM12}, we performed these fits in the radial interval $(r_1 , r_2 )$ such that 
$N_{\rm bh}(< r_1) = 5$  and $N_{\rm bh}(< r_2) = 25$, 
and assume a density profile of constant power-law index.
In  Figure~\ref{msclu2} we show the (scaled) cumulative number of BHs
during the post merger simulations of Section~\ref{relax}. 
We find $N_{\rm bh}(< 0.01{\rm pc}) \approx 200$ at $t= 0.1\times T_{r_{\rm infl}}$
and 
$N_{\rm bh}(< 0.01{\rm pc}) \approx 400$
at $t= 0.3\times T_{r_{\rm infl}}$. 
We can directly compare the numbers so obtained 
with estimates from quasi-steady state 
 Fokker Plank models of the GC.
In their  models, \citet{HA06} assumed $f=0.01$,   and found
$N_{\rm bh}(r < 0.01 {\rm pc}) = 150$; \citet{F06} found similar values. 
This is similar to the number of BHs in our model at the end of the 
inspiral simulations but somewhat smaller than the predicted number of BHs at the end of the post merger simulations.

\begin{figure}
\centering
$\begin{array}{ll}
  \includegraphics[width=0.44\textwidth,angle=0]{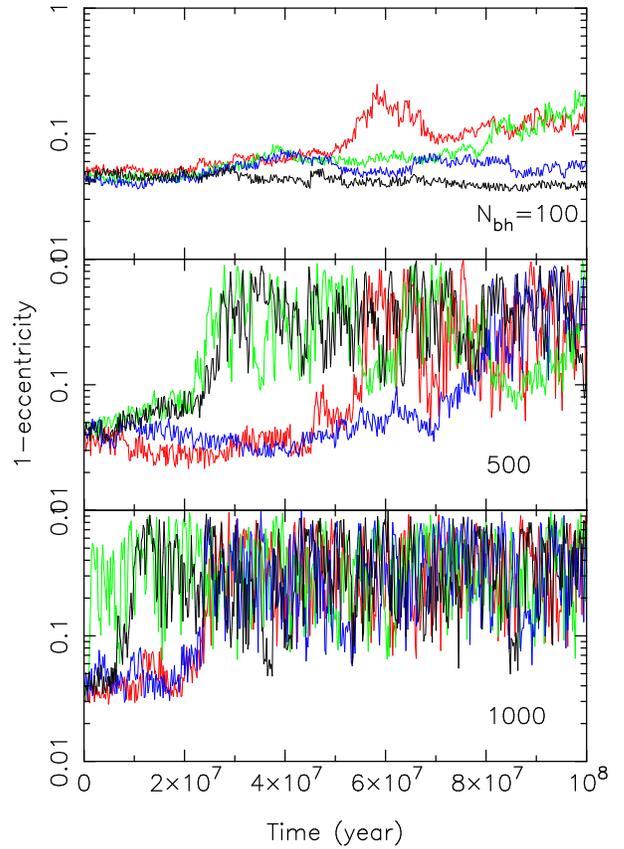}
  \end{array}$
\caption{Eccentricity evolution of test particles taking different numbers 
of $10M_\odot$ perturbers: $N_{0.1}$=100, 500, 1000. 
 The
initial values of the orbital elements were:  a=10~mpc, e=0.95, $\Omega=0$,
$\omega=-\pi/2$ and $i=0.35\pi$. 
The diffusion timescale 
to  evolve to higher angular momenta, where resonant relaxation becomes more efficient 
in randomizing orbital eccentricities,
 decreases the larger the number of BHs in the cusp.
}
  \label{rr-ev}
\end{figure}

 In conclusion, a core in  the density distribution of  stars does not necessarily imply a low density of stellar remnants in the GC.
 In a merger model for NCs, BHs have a smaller core initially -- as a relic of their pre-merger mass 
 segregation, which was not fully erased by the tidal disruption process of the cluster. After a small fraction
 of the relaxation time~($\sim 0.1T_{\rm r_{\rm infl}}$), the BHs dynamically relax due to BH-BH interactions and attain
 a steep central density cusp, while the core in the stellar distribution persists. After this time, 
 the number of BHs inside $0.01~$pc, the radii relevant for EMRIs, is $N_{\rm bh}(<0.01{\rm pc})\sim 100$, 
 a value  comparable to that inferred in steady state models.

\begin{figure}
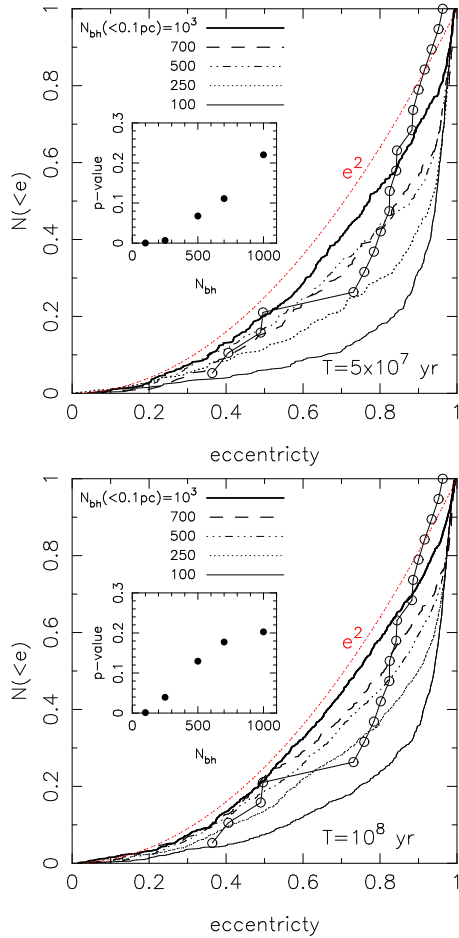

\centering
$\begin{array}{ll}
  \includegraphics[width=0.33\textwidth,angle=0]{Figure21a.ps} \\
    \includegraphics[width=0.33\textwidth,angle=0]{Figure21b.ps} 
  \end{array}$
\caption{Cumulative eccentricity distributions at $5\times 10^7$yr and 
at $10^8$yr of evolution for various values of  the number of $10~M_{\odot}$ BH perutbers inside $0.1$pc, $N_{0.1}$. 
The insert panels give the $p$-value of the K-S tests as a function of $N_{0.1}$. Only when $N_{0.1}\gtrsim 1000$ 
 the eccentricity distribution appears to approach a form that is 
consistent with observations~($p-{\rm value} \gtrsim 0.1$) after $\sim 100~$Myr of evolution (the S-star lifespan). 
}
  \label{s-stars}
\end{figure}

\subsection{Constraining the number of dark remnants at the GC using the S-star orbits}
Observations of the Galctic center have revealed the existence of $\sim 20$ 
young (B-type) stars orbiting  within $\sim 0.05$ pc of the central black hole, the S-stars~\citep{Ghez08,Gil}. The S-stars are 
thought not to have formed in situ where the strong tidal field of the MBH would prevent star formation~\citep{Morriss1993}. 
Mechanisms that invoke the formation of the S-stars farther from the MBH, and subsequent  rapid migration to their
current location  are often invoked  in order to  explain their existence.
The current paradigm  is that the S-stars  result from the  capture of individual 
stars by the tidal disruption of binaries following a  close encounter to the central black hole~\citep{Hills1988}.
This formation model can successfully  accounts for the number of  observed S-stars~\citep{Perets2007}, 
but it results into an orbital distribution which is largely biased toward high eccentricities,
and therefore it is inconsistent with the quasi-thermal eccentricity distribution of the S-star orbits~\citep[e.g.,][]{ant+10}.
Post-migration dynamical evolution due to gravitational perturbations from a field population of BHs
has been invoked in order to bring the predicted orbital distributions more in line with 
observations~\citep{Perets2009,Madigan2011,adrian+14}.
 It has been recently shown that such a scenario could indeed lead to an orbital distribution which reproduces the random 
and eccentric character of the observed orbits~\citep{AM13}.

The timescale over which the S-star orbits evolve and approach the observed distribution 
depends on the number of objects, i.e. BHs, in the stellar cusp,
being the shorter the larger the number  of BHs~\citep{AM13}. 
Accordingly, we  determine  a lower limit to the number of BHs in the GC
by following the dynamical evolution of the S-stars for different values of $N_{\rm bh}$, and
requiring the orbits to approach the observed distribution in  $\sim100~$Myr, 
the  main-sequence life time of a B-type star. 

 We ignore  two-body relaxation since the timescale of interest ($\sim100~$Myr) is short compared with two-body (non-resonant) 
relaxation times near the center of the Milky Way~\citep{M:10}.
Following \citet{AM13} we assume that the semi-major axis distribution, $N(a)$, is known, and it is
 given by the observed values of $a$. The initial orbital eccentricities were assigned randomly
  in the range $0.93\leq e\leq0.99$. 
The orbits were initialized at random times between $0$
and $100$~Myr and followed to a final time of  $100$Myr.  
This setup approximately  reproduces the initial conditions expected in
a binary disruption scenario for the formation of the S-stars~\citep{zhang+13}.
For each S-star (i.e., for each value of a), 100  integrations were carried out 
using the same Hamiltonian model adopted in \citet{MAMW2011} and \citet{AM13}. 
Briefly, the Hamiltonian model accounts for the torques due to finite-$N$ asymmetries in the field-star distribution (resonant-relaxation) 
and 1-PN terms  which result in the (Schwarzschild) precession of the argument of periastron. 
The direction of the torquing field was changed smoothly with time and was randomized in a time of 
 order the precession time of the field population as described in Section VB of~\citet[][]{MAMW2011}.

We consider a field population  of $10~M_\odot$ BHs,which we distributed according to a
 power law density profile $ \rho \sim  r ^{-2}$. The number of 
BHs at radii less than $r$ is
\begin{equation}
N_{\rm bh}(<r)=N_{0.1} \left( r/0.1~{\rm pc}\right),
\end{equation}
where $N_{0.1} $ is the number of BHs within a radius of $0.1$pc; we take
$N_{0.1}=(100,~250,~500,~700,~1000)$, and for each of these values we compared the 
results of our simulations to observations  by performing a two-sample Kolmogorov-Smirnov~(K-S) test on
the eccentricity distributions.

In any of our cusp models,  most orbits have orbital eccentricities that lie initially inside (i.e., at higher eccentricity than)
the Schwarzschild barrier -- the locus of points in the ($E,L$) plane where resonant relaxation is suppressed by fast relativistic precession
of a test particle orbit~\citep{MAMW2011}.
As shown in \citet{AM13},  the Schwarzshild barrier 
behaves as a one-way permeable membrane, which is penetrable  by orbits that approach it from higher eccentricities,
while it is a hard barrier for orbits that approach it from above. 
It can be shown that
the time to diffuse above the Schwarzshild barrier,
towards higher angular momenta,  scales approximately  with the number of perturbers as $T_{d}\sim N^{-3/2}_{\rm bh}(<a)$
or as $\sim N^{-2}_{\rm bh}(<a)$ if the choice for the coherence time is the 
``vector'' resonant relaxation time or the mass precession time
respectively -- i.e., the test particles  diffuse faster toward 
 larger angular momenta, above the Schwarzschild barrier, the larger the number of background perturbers~(Figure~\ref{rr-ev}).

Figure~\ref{s-stars} shows the cumulative eccentricity distributions at $5\times 10^7$yr and 
at $10^8$yr of evolution for various values of  $N_{0.1}$. As the number of BHs in the cusp increases the diffusion  timescale
decreases, allowing the distribution to more rapidly approach 
a quasi-thermal form. The insert panels give
the $p$-value of the K-S tests as a function of $N_{0.1}$. Only for $N_{0.1}\gtrsim 1000$ 
 after $\sim 100~$Myr the eccentricity distribution appears to have approached a form that is 
consistent with observations~($p-{\rm value} \gtrsim 0.1$). These results imply that
 in order to explain the quasi-thermal character of the S-star orbital eccentricities,
the number of BHs within $\sim 0.1~$pc of Sgr~A*  has to be $\gtrsim1000$. 
Interestingly, this number is in agreement  with the predictions of our globular cluster merger model (Figure~\ref{msclu}),
but appears to be much larger than the number of BHs implied by the dynamical models 
 discussed  in Section~(\ref{sec-mass-seg}) which, for $f_{\rm bh}=10^{-3}$, give $N_{0.1} \approx 100$ after $10~$Gyr of evolution.

Interestingly, a globular cluster merger model for the Milky Way NC can potentially 
reconcile models which require high central densities of BHs in order to explain the orbits of the
S-stars, with the  ``missing cusp'' problem of the giant star population.

 \section{Summary}
Understanding the distribution of stellar black holes~(BHs) at the center of the Galaxy
 is fundamental for a variety of astrophysical  problems.
These include the randomization of the S-star orbits, the warping of the young stellar disk, 
and the inspiral of BHs into MBHs -- an important  class of gravitational wave sources
 for the future space-based interferometer antenna eLISA~\citep{AS12}.  
 The efficiency of these processes is very sensitive 
to the number of BHs and to their density distribution near the GC. In this paper we have used
$N$-body simulations to follow the evolution of BHs in models that have not yet reached a collisional
 steady state. Following the evolution of the BHs for timescales of  order the age of the Galaxy, we
made predictions about their density distribution under the two assumptions that:
(1) they follow 
the same phase-space distributions initially as the stars; and (2)  they are initially brought into the Galactic
center by dynamically evolved massive clusters.
Our main results are summarized below.

\begin{itemize}
\item[1)] 
We evolved models that have a parsec-scale density core, and in which the BHs have 
the same phase-space distribution initially as the stars. We found that
the time required for the growth of a relaxed, mass segregated stellar cusp is shorter 
in models which contain a population of heavy remnants~(e.g., Figure~\ref{mass-seg-slp}), and it is sensitive to their
initial number relative to the stars.
\item[2)]  Over the age of the Galaxy, and for a standard IMF, scattering off the BHs has little influence on the evolution of the lighter species.
The time required for the re-growth of a mass segregated stellar cusp can be 
longer than the Hubble time for galaxies similar to the Milky Way~(e.g., Figure~\ref{mass-seg2}). 
\item[3)]  In low and intermediate luminosity galaxies globular clusters can decay to the center of the galaxy through 
dynamical friction and form a compact NC~(Figure~\ref{rvsm-gxs}).
 Such clusters may harbor an inner core cluster of BHs that 
formed and mass-segregated to the center during the cluster evolution. For sufficiently massive 
clusters the BH population is likely to be retained in the cluster center, and then transported to the inner regions
of the galaxy. Thus, massive clusters can represent an efficient  source of BHs  in the central regions of galaxies.
\item[4)]  We used direct $N$-body simulations to follow the inspiral and merger of globular clusters in the GC.
These clusters contained two stellar populations, representing starts and BHs. 
Both standard and top-heavy mass functions were considered.
The BHs were initially segregated
to the cluster center. After about 10 inspiral  events
the  formed NC developes a density profile that falls off as $\sim r ^{-1.5}$, and a shallower core-like 
profile, $\gamma \lesssim 1$, inside a radius $\sim 0.5$pc. These properties are 
similar to those observed in the Milky Way NC. 
We find that the initial mass-segregation is not completely erased as the clusters are disrupted by the MBH tidal 
field~(Figure~\ref{density-cont}).
As a consequence of this,   in the merger end-product the BHs dominate the total mass density   within
a radius of approximately $0.3~$pc~(Figure~\ref{DEN1}).
\item[5)]  By continuing the evolution of the model after the final inspiral event, we find that 
  the BH population rapidly relaxes and attains a steep central density cusp, $r^{-2.2}$.
By half a relaxation time, or roughly $10$~Gyr~(when scaled to the Milky Way), the 
 core that was formed in the stellar distribution shrinks to $0.1-0.2~$pc~(Figure~\ref{msclu}).  While
 the density profile slope outside these radii remained nearly unchanged  the densities
decreased as a consequence of heating of the stars by the BHs.  
Gravitational encounters also caused the NC to evolve toward spherical symmetry in configuration and velocity space.
 \item[6)] We studied the orbital evolution of the S-stars under the assumption that they were deposited to
 the GC through the disruption of binary stars by the MBH.
 Our analysis included the joint effects of Newtonian and relativistic perturbations to the motion, including 
 the torques due to finite-$N$ asymmetries in the field-star distribution~(resonant relaxation).
 We evolved the S-star orbits for a time of order $10^8$yr adopting models for the GC characterized by 
different number densities  of BHs.
 We find that in order for the S-stars  to achieve a nearly thermal distribution of eccentricities during their 
 lifetime the GC should contain $\gap 1000$ 
 BHs inside $0.1~$pc~(Figure~\ref{s-stars}). We argue that  this lower limit for the number of BHs
 at the GC  is consistent with a dissipationless formation model
 for the origin of the Milky Way NC.
\end{itemize}

I am grateful to R. Capuzzo-Dolcetta, D.~Merritt, A. Mstrobuono-Battisti for numerous discussions about the 
ideas presented in this paper. I thank D.~Merritt and H.~Perets for useful comments that helped to improve 
an earlier version of this manuscript.

This research was partially supported by 
the National Aeronautics and Space Administration under grant no. NNX13AG92G. 
Computations were performed on Sunnyvale computing cluster
at CITA and  also on the ARC supercomputer at the SciNet HPC Consortium. SciNet is funded by: the
Canada Foundation for Innovation under the auspices of Compute
Canada; the Government of Ontario; Ontario Research FundResearch Excellence; and the University of Toronto.


\begin{thebibliography}{}
\bibitem[Alexander~(2005)]{AX:05}Alexander, T. \ 2005, Phys. Rep., 419, 65
\bibitem[Alexander \& Hopman~(2009)]{AH:09} Alexander, T., \& Hopman, C. 2009, ApJ, 697, 1861
\bibitem[Amaro-Seoane et al.~(2012)]{AS12} Amaro-Seoane, P., et al. \ 2012, Classical and Quantum Gravity, 29, 124016 
\bibitem[Amaro-Seoane \& Xian~(2013)]{amaro-xian13}Amaro-Seoane, P., and Xian, C. \ 2013, arXiv:1310.0458 
\bibitem[Antonini et al.~(2009)]{ACM09}Antonini, F.,  Capuzzo-Dolcetta, R., \& Merritt, D., 2009, MNRAS, 399, 671
\bibitem[{{Antonini} {et~al.}(2010){Antonini}, {Faber}, {Gualandris}, \& {Merritt}}]{ant+10}
{Antonini}, F., {Faber}, J., {Gualandris}, A., \& {Merritt}, D. 2010, \apj,
 713, 90
\bibitem[Antonini \& Merritt~(2012)]{AM12}Antonini, F. \& Merritt, D. \ 2012, ApJ, 745, 83
\bibitem[Antonini \& Merritt~(2013)]{AM13}Antonini, F.  \& Merritt, D. \ 2013, ApJ, 763, L10
\bibitem[Antonini~et~al.~(2012)]{AM12-2}Antonini, F., Capuzzo-Dolcetta, R., Mastrobuono-Battisti, A. \& Merritt, D. \ 2012, ApJ, 750, 111 
\bibitem[Antonini~(2013)]{me13}Antonini, F. \ 2013, ApJ, 763, 62
\bibitem[Antonini \& Perets~(2012)]{me+perets12}Antonini, F. \& Perets, H. 2012, ApJ, 757, 27


\bibitem[Bahcall \& Wolf~(1976)]{BW:76} Bahcall, J.~N., \& Wolf, R.~A.\ 1976, \apj, 209, 214
\bibitem[Bailey \& Davies~(1999)]{bailey+davies99}Bailey, V. C., \& Davies, M. B. 1999, MNRAS, 308, 257
\bibitem[Bartko et al.~(2010)]{B:10} Bartko et al.  \ 2010, \apj, 708, 834
\bibitem[Banerjee \& Kroupa~(2011)]{BK11}Banerjee, S., \& Kroupa, P. \ 2011, ApJ, 741, L12
\bibitem[Banerjee et al. (2010)]{BBK10}Banerjee, S.,Baumgardt, H., Kroupa, P. \ 2010, MNRAS, 402, 371
\bibitem[B\"{o}ker et al.~(2002)]{BSM:02} B\"{o}ker, Laine, S., van der Marel,
R.~P., Sarzi, M., Rix, H., Ho, L.~C., \& Shields, J.~C. \ 2002, AJ, 123, 1389
\bibitem[Brassington et al.~(2010)]{brassington+10}Brassington, N. J. et al. \ 2010, ApJ, 725, 1805
\bibitem[Buchholz et al.~(2009)]{BSE}Buchholz, R.~M., Sch\"{o}del, R., \& Eckart, A.  2009, A\&A, 499, 483

\bibitem[Carlberg \& Hartwick~(2014)]{carlberg+hartwick14}Carlberg, R., Hartwick, F.~D.~A. \ 2014, arXiv1401.4742
\bibitem[Carollo et al.~(1998)]{carollo} Carollo, C. M., Stiavelli, \& Mack, J. \ 1998, AJ, 116, 68
\bibitem[Capuzzo-Dolcetta \& Miocchi~(2008)]{CDM08} Capuzzo-Dolcetta, R., Miocchi, P., 2008, MNRAS, 388, L69
\bibitem[Capuzzo-Dolcetta~(1993)]{capuzzo93}Capuzzo-Dolcetta, R. \ 1993, ApJ, 415, 616
\bibitem[Casertano et al.~(1987)]{CPV} Casertano, S., Phinney, E. S., \& Villumsen, J. V., 1987, IAU Symp. 127,
Structure and Dynamics of Elliptical Galaxies, ed. T. de Zeeuw (Dordrecht: Reidel), 475
\bibitem[C\^ot\'{e} et al.~(2006)]{Cote} C\^ot\'{e} et al. 2006 ApJS, 165, 57

\bibitem[Dale et al.~(2009)]{dale+09} Dale, J.~E., Davies, M.~B., Church, R.~P., \& Freitag, M. \ 2009, \mnras, 393, 1016
\bibitem[Davies \& King~(2005)]{davies+king05}Davies, M. B., \& King, A. \ 2005, ApJ, 624, L25
\bibitem[De~Lorenzi~et~al.~(2013)]{de-lorenzi+13}De Lorenzi, F., Hartmann, M.; Debattista, V. P., Seth, A. C., \& Gerhard, O. \ 2013, MNRAS, 429, 2974
\bibitem[Do et al.~(2009)]{D:09}Do, T., Ghez, A.~M., Morris, M.~R., Lu, J.~R., Matthews, K., Yelda, S., \& Larkin, J. \  2009, ApJ, 703, 1323
\bibitem[Do et al.~(2013)]{do+13}Do, T. et al. \ 2013, ApJ, 779L, 6
\bibitem[Dowing et al.~(2010)]{DBGS10}Downing, J. M. B., Benacquista, M. J., Giersz, M., \& Spurzem, R. \ 2010, MNRAS, 407, 1946
\bibitem[Dowing et al.~(2011)]{DBGS11}Downing, J. M. B., Benacquista, M. J., Giersz, M., \& Spurzem, R. \ 2011, MNRAS, 416, 133

\bibitem[Feldmeier et al.~(2014)]{Feldme+14} Feldmeier, A. et al.\ 2014  arXiv:1406.2849
\bibitem[Ferrarese et al.~(2006)]{F:06} Ferrarese, L. et al. \ 2006, ApJ, 644, L21
\bibitem[Figer et al.~(2004)]{figer+04}Figer, D. F., Rich, R. M., Kim, S. S., Morris, M., \& Serabyn, E. \ 2004, ApJ, 601, 319
\bibitem[Freitag et al.~(2006)]{F06}Freitag, M., Amaro-Seoane, P., \& Kalogera, V. \ 2006, ApJ, 649, 91
\bibitem[Forbes et al.~(2008)]{FORBES}Forbes, D. A., Lasky, P., Graham, A. W., \& Spitler, L. 2008, MNRAS, 389, 1924

\bibitem[Gaburov et. al.~(2009)]{sap}Gaburov, E., Harfst,  S., \& Portegies Zwart, S.  \
\bibitem[Genzel et al.~(2003)]{Genzel03}Genzel, R. et al. \ 2003, ApJ, 594, 812
\bibitem[Ghez et al.~(2008)]{Ghez08} Ghez, A. M. et al., 2008, ApJ, 689, 1044
\bibitem[Gillessen~(2009)]{Gil} Gillessen, S., Eisenhauer, F., Trippe, S., Alexander, T., Genzel, R., Martins, F., Ott, T. 2009, ApJ, 692, 1075
\bibitem[{{Genzel} {et~al.}(2010){Genzel}, {Eisenhauer}, \&
 {Gillessen}}]{gen+10}
{Genzel}, R., {Eisenhauer}, F., \& {Gillessen}, S. 2010, Reviews of Modern
 Physics, 82, 3121 
 \bibitem[Gnedin et al.~(2013)]{gnedin+13} Gnedin, O.~Y., Ostriker, J.~P., \& Tremaine~S. \ 2013, arXiv1308.0021
\bibitem[Graham \& Spitler~(2009)]{GS:09} Graham, A. W. \& Spitler, L. R. 2009, MNRAS, 397, 2148
\bibitem[Gualandris \& Merritt (2012)]{GM12}Gualandris, A. \& Merritt, D. 2012, ApJ, 744, 74
\bibitem[G\"{u}rkan et al.~(2005)]{gurkan+05}G\"{u}rkan, M. A., \& Rasio, F.~A. \ 2005, ApJ, 628, 236
 
\bibitem[Haller et al.~(1996)]{haller+96}Haller, J. W., Rieke, M. J., Rieke, G. H., et al. \ 1996, ApJ, 456, 194
\bibitem[Hamers et al.~(2014)]{adrian+14} Hamers, A., Portegies, Zwart, S.~F. \& Merritt, D. \ 2014, in preparation
\bibitem[Hashimoto et al.~(2003)]{Hashimoto} Hashimoto, Y., Funato, Yoko, \& Makino, J. \ 2003, ApJ, 582, 196
\bibitem[Harfst et al.~(2006)]{HGMM0}Harfst, S., Gualandris, A., Merritt, D., Spurzem, R., Portegies Zwart, S., Berczik, P. \ 2006,
NewA, 12, 357
\bibitem[Harfst et al. (2008)]{HGMM} Harfst, S., Gualandris, A., Merritt, D., Mikkola, S. \ 2008, MNRAS, 389, 2
\bibitem[Hartmann et al.~(2011)]{Hartmann} Hartmann, H., Debattista,  V.~P., Seth, A., Cappellari, M., \& Quinn, T.~R., \ 2011, arxiv:1103.5464
\bibitem[Heggie \& Hut~(2003)]{HH03} Heggie, D., Hut, P. \ 2003, gmbpbook
\bibitem[Hills~(1998)]{Hills1988}Hills, J.~G. \ 1988, Nature, 331, 687
\bibitem[Hopman \& Alexander~(2006)]{HA06} Hopman, C., \& Alexander, T., 2006, ApJL, 645, 133
\bibitem[Hopman \& Alexander~(2005)]{HA05} Hopman, C., \& Alexander, T., 2005, ApJ, 629, 362

\bibitem[\protect\citeauthoryear{Katz}{1991}]{Katz:1991} Katz, N., 1991, ApJ, 368, 325
\bibitem[King~(1962)]{K62} King, I. R., 1962, AJ, 67, 471
\bibitem[Kocsis \& Tremaine~(2011)]{kocsis+tremaine11}Kocsis, B., \& Tremaine, S. \ 2011, MNRAS, 412, 187

\bibitem[Launhardt et al.~(2002)]{LZM} Launhardt, R., Zylka, R., \& Mezger, P. G. 2002, A\&A, 384, 112
\bibitem[Leigh~et al.~(2012)]{Nathan}Leigh, N., B\"{o}ker, T., Knigge, C. \ 	2012, MNRAS, 424, 2130
\bibitem[L\"{o}ckmann et al.~(2010)]{lockmann+10} L\"{o}ckmann, U., Baumgardt, H., \& Kroupa, P. \ 2010, MNRAS, 402, 519

\bibitem[Maccarone et al.~(2007)]{maccarone+07} Maccarone, T. J., Kundu, A., Zepf, S. E., \& Rhode, K. L. \  2007, Nature, 445, 183
\bibitem[Maccarone et al.~(2011)]{maccarone+11} Maccarone, T. J., Kundu, A., Zepf, S. E., \& Rhode, K. L. \  2011, MNRAS, 410, 1655
\bibitem[Madigan et al. (2011)]{Madigan2011}Madigan, A., Hopman, C., \&Levin, Y. \ 2011, ApJ, 738, 99

\bibitem[Makino \& Taiji~(1998)]{MakinoGRAPE} 
Makino, J., \& Taiji, M.\ 1998, Scientific Simulations with Special-Purpose Computers--the GRAPE Systems, by Junichiro Makino, Makoto Taiji, pp.~248.~ISBN 0-471-96946-X.~Wiley-VCH , April 1998


\bibitem[Mastrobuono-Battisti \& Perets~(2013)]{mp13}Mastrobuono-Battisti, A., \& Perets, H.~B. \ 2013, ApJ, 779, 85
\bibitem[Mastrobuono-Battisti \& Perets~(2014)]{mpl14}Mastrobuono-Battisti, A., Perets, H.~B., Loeb, A. \ 2014, arXiv1403.3094

\bibitem[Merritt~(2009)]{M09}Merritt, D.~2009, ApJ, 694, 959
\bibitem[Merritt~(2010)]{M:10} Merritt, D. \ 2010, ApJ, 718, 739
\bibitem[Merritt \& Vasiliev~(2011)] {MV2011}Merritt, D., \& Vasiliev, E. \ 2011, ApJ, 726, 61
\bibitem[Merritt et al.~(2011)]{MAMW2011}Merritt, D., Alexander, T., Mikkola, S., \& Will, C. \ 2011,PhRvD, 84, 4024
\bibitem[Merritt~(2013)]{Merritt-book}Merritt, D., Dynamics and Evolution of Galactic Nuclei, 2013, Princeton University Press
\bibitem[Milosavljevi\'{c}~(2004)]{Milos04} Milosavljevi\'{c}, M. \ 2004, ApJ, 605, L13
\bibitem[Morscher et al.~(2013)]{morscher+13}Morscher, M., Umbreit, S., Farr, W.~M., \& Rasio, F.~A. \ 2013, ApJ, 763, L15
\bibitem[Morris~(1993)]{Morriss1993}Morris, M. 1993, ApJ, 408, 496
\bibitem[Muno et al.~(2005)]{muno+05}Muno, M. P., Pfahl, E., Baganoff, F. K., Brandt, W. N., Ghez, A., Lu, J., \& Morris, M. R. 2005ApJ, 622, L113
\bibitem[Neumayer \& Walcher~(2012)]{NW12}Neumayer, N., \& Walcher, C. J. 2012, Adv. Astron., 2012, 709038

\bibitem[Oh et al.~(2009)]{okf}
Oh, S., Kim, Sungsoo, S., \& Figer, D.~F. \ 2009, JKAS, 42, 17
\bibitem[Pfuhl et al.~(2011)]{Ol11}Pfuhl, O. et al. \ 2011, ApJ, 741, 108

\bibitem[Perets et  al.~(2007)]{Perets2007}Perets, H.~B., Hopman, C., \& Alexander, T.  \ 2007, ApJ, 656, 709 
\bibitem[Perets et al. (2009)]{Perets2009}Perets, H.~B.,  Gualandris, A.,  Kupi, G., Merritt, D., \& Alexander, T. \ 2009,  ApJ, 702, 884
\bibitem[Perets \& Mastrobuono-Battisti~(2014)]{hagai+ale} Perets, H. \& Mastrobuono Battisti, A. \ 2014, ArXiv 
\bibitem[Poon \& Merritt~(2004)]{PM:04}Poon, M.~Y., \& Merritt, D. \ 2004, ApJ, 606, 774
\bibitem[Preto~et al.~(2004)]{PR04}Preto, M., Merritt, D., Spurzem, R. \ 2004, ApJ, 613, L109
\bibitem[Preto \& Amaro-Seoane~(2010)]{preto-amaro10}Preto, M., \& Amaro-Seoane, P. 2010, ApJL, 708, L42
\bibitem[Rayleigh~(1919)]{rl-19} Rayleigh, L. \ 1919, Phil. Mag., 37, 321
\bibitem[Rossa~et al.~(2006)]{Rossa}Rossa, J. et al. 2006, AJ, 132, 1074
\bibitem[Seth et al.~(2008)]{seth} Seth, A. et al. \ 2008, ApJ, 678, 116
\bibitem[Schinnerer et al.~(2008)]{Sch08} Schinnerer E., B\"{o}ker T.,  Meier D. S., \& Calzetti D. \ 2008, ApJL, 684, L21
\bibitem[Sch\"odel et al.~(2002)]{sch02}Sch\"odel et al. \ 2002, Nature, 419, 694
\bibitem[Sch\"{o}del et al.~(2009)]{S09} Sch\"{o}del, R., Merritt, D., \& Eckart \ 2009 A\&A, 502, 91
\bibitem[Sch\"{o}del et al.~(2014)]{S14}Sch\"{o}del et al. 2014, A\&A, 566, 47
\bibitem[Scott \& Graham~(2012)]{SG12}Scott, N., \& Graham, A.~W. \ 2012, arXiv1205.5338
\bibitem[Sippe \& Hurley~(2013)]{sippe+hurley13}Sippel, A.~C. \& Hurley, J.~R. \ 2013, \mnras, 430, L30
\bibitem[Spitzer~(1987)]{Spitzer}Spitzer, L. \ 1987, Dynamical evolution of Globular Clusters (Princeton: Princeton Univ. Press)
\bibitem[Strader et al.~(2012)]{strader+12} Strader, J., Chomiuk, L., Maccarone, T. J., Miller-Jones,
J. C. A., \& Seth, A. C. 2012, Nature, 490, 71
\bibitem[Tremaine et al.~(1975)]{TOS} Tremaine, S. D., Ostriker, J. P., \& Spitzer, L., Jr. 1975, ApJ, 196, 407
\bibitem[Trippe et al.~(2008)]{trippe}Trippe et al. \ 2008, A\&A, 492, 419
\bibitem[Turner et al.~(2012)]{turner+12}Turner, M. et al. \ 2012, ApJS, 203, 5
\bibitem[Wehner  \& Harris~(2006)]{WH:06}Wehner, E. H. \& Harris, W. E. 2006, ApJ, 644, L17
\bibitem[Yusef-Zadeh et al.~(2012)]{yz+12}Yusef-Zadeh, F., Bushouse, H., \& Wardle, \ 2012, ApJ, 744, 24
\bibitem[Zhang et al.~(2013)]{zhang+13}Zhang, Fupeng, Lu, Youjun, \& Yu, Qingjuan \ 2013, ApJ, 768, 153
\end{thebibliography}
\end{document}